\documentclass[10pt,aps,twocolumn,secnumarabic,balancelastpage,amsmath,amssymb,nofootinbib,hyperref=pdftex,prx,citeautoscript,showpacs]{revtex4-1}

\usepackage{graphics}
\usepackage[pdftex]{graphicx}
\usepackage{tabularx}
\graphicspath{{./Figures/}}
\usepackage{longtable}
\usepackage{amsmath}
\usepackage{color,soul}
\usepackage{braket}
\usepackage{float}

\makeatletter
\def\p@subsection{\thesection\,}
\makeatother

\newcommand{\bo}[1]{{\bf #1}}

\newcommand{\kk}{{\bo k}}

\usepackage{hyperref}
\hypersetup{colorlinks=true,linkcolor=blue,anchorcolor=blue,citecolor=blue,filecolor=blue,urlcolor=blue,bookmarksnumbered=true,pdfview=FitB}

\begin{document}

\title{Proximity effect in a two-dimensional electron gas coupled to a thin superconducting layer}
\author{Christopher Reeg}
\author{Daniel Loss}
\author{Jelena Klinovaja}
\affiliation{Department of Physics, University of Basel, Klingelbergstrasse 82, CH-4056 Basel, Switzerland}
\date{\today}
\begin{abstract}
There have recently been several experiments studying induced superconductivity in semiconducting two-dimensional electron gases that are strongly coupled to thin superconducting layers, as well as probing possible topological phases supporting Majorana bound states in such setups. We show that a large band shift is induced in the semiconductor by the superconductor in this geometry, thus making it challenging to realize a topological phase. Additionally, we show that while increasing the thickness of the superconducting layer reduces the magnitude of the band shift, it also leads to a more significant renormalization of the semiconducting material parameters and does not reduce the challenge of tuning into a topological phase.
\end{abstract}

\maketitle

\section{Introduction} Topological superconductors host zero-energy Majorana bound states at their edges that are highly sought for applications in topological quantum computing \cite{Kitaev:2001,Nayak:2008,Alicea:2012}. The two proposals to realize topological superconductivity that have received the most attention to date involve engineering Majorana bound states in either low-dimensional semiconducting systems \cite{Sato:2009,Sato:2009b,Lutchyn:2010,Oreg:2010,Sau:2010,Alicea:2010,Chevallier:2012,Halperin:2012,Sticlet:2012,Klinovaja:2012,Klinovaja:2012c,Prada:2012,Dominguez:2012,Klinovaja:2013b,DeGottardi:2013,Maier:2014,Vernek:2014,Weithofer:2014,Thakurathi:2015,Dmytruk:2015} or in ferromagnetic atomic chains \cite{NadjPerge:2014,Ruby:2015,Pawlak:2016,Klinovaja:2013,Vazifeh:2013,Braunecker:2013,NadjPerge:2013,Pientka:2013,Awoga:2017}. After the first signatures of topological superconductivity were observed \cite{Mourik:2012,Deng:2012,Das:2012,Churchill:2013,Finck:2013}, much of the experimental focus was placed on developing more suitable devices for realizing robust topological superconducting phases. One of the most significant experimental advances of the past few years was the successful epitaxial growth of thin layers of superconducting Al on InAs and InSb nanowires \cite{Chang:2015,Albrecht:2016,Deng:2016,Gazibegovic:2017,Zhang:2017_2}. The intimate contact between the semiconductor and superconductor in these devices ensures a hard induced superconducting gap. Recently, this epitaxial growth technique has been applied also to InAs two-dimensional electron gases (2DEGs) \cite{Kjaergaard:2016,Shabani:2016,Kjaergaard:2017,Suominen:2017,Nichele:2017}.

The proximity effect has been theoretically studied recently in both strict-one-dimensional (1D) \cite{Reeg:2017_3} and quasi-1D \cite{Reeg:2018} wires coupled to thin superconducting layers. In both instances, a strong proximity coupling induces a large band shift on the semiconducting wire that is comparable to the level spacing in the superconductor, $\delta E_s=\pi\hbar v_F/d$ (which is $\delta E_s\sim400$ meV for superconductor thickness $d\sim10$ nm and Fermi velocity of Al $v_F\sim2\times10^6$ m/s). In both cases, this large band shift makes it very challenging to realize a topological phase when utilizing thin superconducting layers.

In this paper, we extend the works of Refs.~\cite{Reeg:2017_3,Reeg:2018} to the 2D limit. We show that the large band shift that plagues the 1D case persists also in 2D. First, we show that the self-energy induced in an infinite 2DEG coupled to a superconductor of finite thickness is equivalent to that induced in an infinite wire coupled to a 2D superconductor of finite width (corresponding to the theoretical model of Ref.~\cite{Reeg:2017_3}), with the simple replacement of a 1D momentum by the magnitude of a 2D momentum. Analyzing the self-energy, we find that the induced gap in the presence of only Rashba spin-orbit coupling can be made comparable to the bulk gap of the superconductor only if the tunneling energy scale exceeds the large level spacing of the superconducting layer. As in the 1D case, the large tunneling energy scale induces a large band shift on the 2DEG and makes it very challenging to realize a topological phase. We also show that while the band shift can be significantly reduced by increasing the thickness of the superconducting layer, the topological phase is still difficult to realize if the 2DEG/superconductor interface remains very transparent.

\begin{figure}[t!]
\centering
\includegraphics[width=\linewidth]{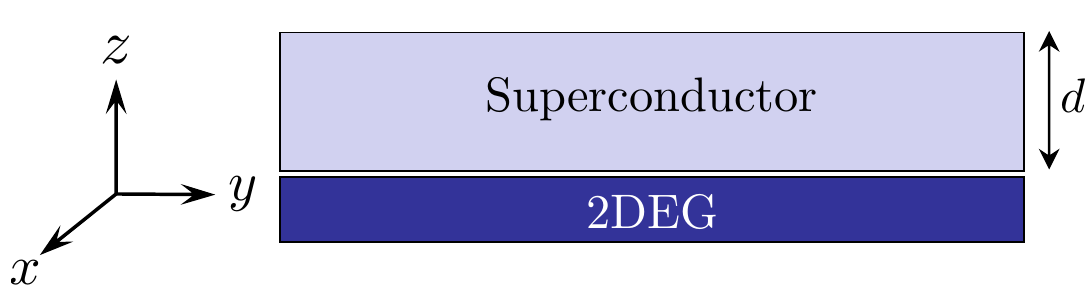}
\caption{\label{setup} A 2DEG is proximity-coupled to an $s$-wave superconductor with finite thickness $d$. Both systems are taken to be infinite in the $xy$-plane.}
\end{figure}


\section{Model of the Proximity Effect} \label{Sec2}
The system we consider  consists of a 2DEG with strong Rashba spin-orbit interaction (SOI) proximity-coupled to an $s$-wave superconductor of thickness $d$, as shown in Fig.~\ref{setup}. The 2DEG-superconductor heterostructure is described by the action
\begin{equation} \label{Hamiltonian}
S=S_{2D}+S_{s}+S_t.
\end{equation}
The action of the 2DEG in Nambu space is given by
\begin{equation} \label{H2D}
S_{2D}=\frac{1}{2}\int_{\kk,\omega}c^\dagger_{\kk,\omega}\left(i\omega-\mathcal{H}^{2D}_\kk\right)c_{\kk,\omega},
\end{equation}
where $\omega$ is a Matsubara frequency, $\kk=(k_x,k_y)$ is the momentum, $\int_{\kk,\omega}\equiv\int d\omega/2\pi\int d\kk/(2\pi)^2$, and $c_{\kk,\omega}=(c_{\kk,\omega,\uparrow},c_{\kk,\omega,\downarrow},c^\dagger_{-\kk,-\omega,\uparrow},c^\dagger_{-\kk,-\omega,\downarrow})^T$ is a spinor of Heisenberg operators describing states in the 2DEG. The Hamiltonian density is
\begin{equation}
\mathcal{H}_\kk^{2D}=\xi_k\tau_z+\alpha(k_y\sigma_x-k_x\tau_z\sigma_y),
\end{equation}
where $\xi_k=k^2/2m_{2D}-\mu_{2D}$ ($m_{2D}$ and $\mu_{2D}$ are the effective mass and chemical potential of the 2DEG, respectively, and  $k^2=k_x^2+k_y^2$), $\alpha$ is the Rashba SOI constant, and $\sigma_{x,y,z}$ ($\tau_{x,y,z}$) are Pauli matrices acting in spin (Nambu) space. The superconductor is described by the BCS action,
\begin{equation} \label{Hs}
\begin{aligned}
S_{s}=\frac{1}{2}\int_{\kk,\omega}\int_0^d dz\,\eta^\dagger_{\kk,\omega}(z)[i\omega-\mathcal{H}_k^s(z)]\eta_{\kk,\omega}(z),
\end{aligned}
\end{equation}
where $\eta_{\kk,\omega}=[\eta_{\kk,\omega,\uparrow},\eta_{\kk,\omega,\downarrow},\eta_{-\kk,-\omega,\uparrow}^\dagger,\eta^\dagger_{-\kk,-\omega,\downarrow}]^T$ is a spinor of Heisenberg operators describing states in the superconductor and the Hamiltonian density is
\begin{equation}
\mathcal{H}^{s}_k(z)=\left(-\frac{\partial_z^2}{2m_s}+\frac{k^2}{2m_s}-\mu_s\right)\tau_z-\Delta\sigma_y\tau_y,
\end{equation}
with $m_s$, $\mu_s$, and $\Delta$ the effective mass, chemical potential, and pairing potential of the superconductor, respectively. Local tunneling at the interface between the two materials is assumed to conserve both spin and momentum,
\begin{equation} \label{St}
S_t=-\frac{t}{2}\int_{\kk,\omega}[\eta_{\kk,\omega}^\dagger(z_{2D})\tau_zc_{\kk,\omega}+H.c.],
\end{equation}
where $t$ is the tunneling amplitude. We must take the 2DEG to be located at some finite $z_{2D}$ ($0<z_{2D}<d$) due to the breakdown of the tunneling Hamiltonian approach for the case where the 2DEG is located at the boundary of the superconductor. The breakdown of the tunneling Hamiltonian results from our neglect of the thickness of the 2DEG (for related calculations in which the finite thickness is taken into account, see Refs.~\cite{Volkov:1995,Fagas:2005,Tkachov:2005,Reeg:2016}). However, as shown in Ref.~\cite{Reeg:2017_3}, choosing $k_Fz_{2D}\ll1$ (where $k_F=\sqrt{2m_s\mu_s}$ is the Fermi momentum of the superconductor) yields good agreement with numerical calculations in which there is no issue with placing the 2DEG strictly at the boundary.

In the absence of tunneling, the spectrum of the 2DEG consists of two spin-orbit-split subbands described by
\begin{equation} \label{bareRashba}
E^2_\pm(k)=(\xi_k\pm\alpha k)^2.
\end{equation}
When the finite-size quantization scale of the superconductor greatly exceeds the gap, $1/m_sd^2\gg\Delta$, the first few subbands of the superconductor follow a linearized form given by ($\hbar=1$)
\begin{equation} \label{bareSC}
E_n^2(k)=\left([k_Fd/\pi-n]\delta E_s-\frac{k^2}{2m_s}\right)^2+\Delta^2,
\end{equation}
where $\delta E_s=\pi v_F/d$ is the level spacing in the superconductor ($v_F=k_F/m_s$ is the Fermi velocity) and $n\in\mathbb{Z}^+$. When the thickness of the superconducting layer is much smaller than its coherence length, $d\ll\xi_s=\pi v_F/\Delta$, the level spacing of the layer greatly exceeds its gap, $\delta E_s\gg\Delta$. 

\begin{figure}[b!]
\centering
\includegraphics[width=.8\linewidth]{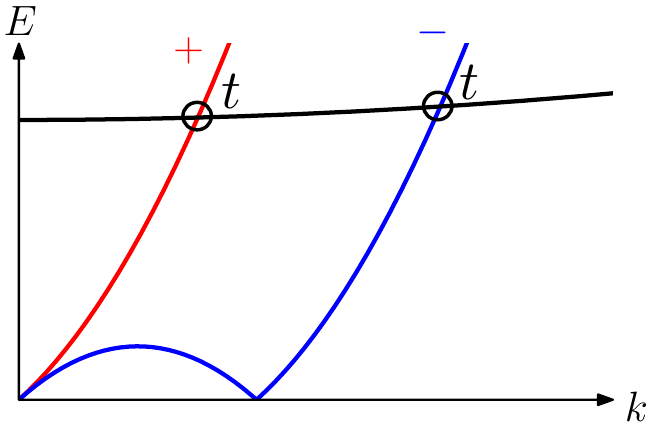}
\caption{\label{barespectra} Sketch of Bogoliubov excitation spectrum as a function of $k=\sqrt{k_x^2+k_y^2}$ in the absence of tunneling, assuming $\delta E_s\gg E_{so}$ and $\mu_n=0$. Red and blue curves correspond to $\pm$ subbands of 2DEG [Eq.~\eqref{bareRashba}], respectively, which result from the spin-splitting Rashba SOI. The black curve corresponds to lowest-energy subband of superconductor [Eq.~\eqref{bareSC}]. A weak tunneling amplitude $t$ induces anticrossings in the spectrum where indicated and induces a superconducting gap in the 2DEG at the Fermi momenta [corresponding to those momenta for which $E_\pm(k)=0$]. Due to the large energy mismatch between the superconducting subband and the Fermi points of the 2DEG, the induced gap is very small.} 
\end{figure}

The spectrum of both the 2DEG and the superconductor are plotted in Fig.~\ref{barespectra}. Provided that $\min(|k_Fd/\pi- n|)\gg\Delta/\delta E_s$, the bands of the 2DEG and superconductor intersect at high energies $E\sim\delta E_s\gg\Delta$. Since we impose momentum conservation (in addition to energy conservation) in {Eq.~\eqref{St}}, the subbands are coupled only at the intersection points. Thus, a weak tunnel coupling induces anticrossings in the spectrum, as indicated in Fig.~\ref{barespectra}, that leads to a shift in the subbands of the 2DEG. Additionally, the tunnel coupling opens a superconducting gap at the Fermi momenta of the 2DEG; however, due to the intersection points lying at very large energies, the gap opened in the 2DEG is very small. A large gap can only be induced if tunneling is strong enough to overcome the large energy mismatch $\sim\delta E_s$.

To determine the self-energy of the 2DEG induced by the superconductor, we integrate out the superconducting degrees of freedom. After integrating out, the 2DEG can be described by the effective action
\begin{equation} \label{Seff}
S_\text{eff}=\int_{\kk,\omega}c^\dagger_{\kk,\omega}(i\omega-\mathcal{H}_\kk^{2D}-\Sigma_{k,\omega})c_{\kk,\omega},
\end{equation}
with the self-energy given by
\begin{equation} \label{selfenergy}
\Sigma_{k,\omega}=t^2\tau_zG^s_{k,\omega}(z_{2D},z_{2D})\tau_z.
\end{equation}
In Eq.~\eqref{selfenergy}, $G^s_{k,\omega}(z,z')$ is the Green's function of the bare superconductor (in the absence of tunneling), which satisfies
\begin{equation} \label{GorkovEq}
[i\omega-\mathcal{H}_k^s(z)]G_{k,\omega}^s(z,z')=\delta(z-z').
\end{equation}
Imposing a vanishing boundary condition at $z=0$ and $z=d$, we find a solution to Eq.~\eqref{GorkovEq} given by
\begin{widetext}
\begin{equation} \label{bareG}
\begin{aligned}
G_{k,\omega}^s(z,z')&=\frac{1}{2v_F\Omega}(i\omega-\Delta\tau_y\sigma_y+i\Omega\tau_z)\left\{\frac{\sin[k_+(d-z')]}{\sin(k_+d)}e^{ik_+z}+[i+\cot(k_+d)]\sin(k_+z')e^{-ik_+z}\right\} \\
	&+\frac{1}{2v_F\Omega}(i\omega-\Delta\tau_y\sigma_y-i\Omega\tau_z)\left\{[-i+\cot(k_-d)]\sin(k_-z')e^{ik_-z}+\frac{\sin[k_-(d-z')]}{\sin(k_-d)}e^{-ik_-z}\right\}+G_{k,\omega}^\text{bulk}(z-z'),
\end{aligned}
\end{equation}
where $k_\pm^2=2m_s(\mu_s\pm i\Omega)-k^2$ and $\Omega^2=\Delta^2+\omega^2$ \cite{Reeg:2017,Reeg:2017_3}. The Green's function of a bulk superconductor, expressed in real space, is
\begin{equation} \label{bulkG}
\begin{aligned}
G_{k,\omega}^\text{bulk}(z-z')&=-\int\frac{dk_z}{2\pi}\frac{i\omega+(\xi_k+k_z^2/2m_s)\tau_z-\Delta\tau_y\sigma_y}{\omega^2+(\xi_k+k_z^2/2m_s)^2+\Delta^2}e^{ik_z(z-z')} \\
	&=-\frac{1}{v_F\Omega\varphi}\biggl[(i\omega-\Delta\tau_y\sigma_y)\cos(\zeta|z-z'|)-\Omega\tau_z\sin(\zeta|z-z'|)\biggr]e^{-\chi|z-z'|},
\end{aligned}
\end{equation}
where, in evaluating the integral, we make a semiclassical expansion $k_\pm=k_F\varphi\pm i\Omega/(v_F\varphi)\equiv\zeta\pm i\chi$ (valid in the limit $\mu_s\gg\Omega$) and define a quantity $\varphi^2=1-k^2/k_F^2$ that parametrizes the trajectories of states in the superconductor. Substituting the Green's function Eq.~\eqref{bareG} into the self-energy Eq.~\eqref{selfenergy}, we find
\begin{equation} \label{selfenergy2}
\Sigma_{k,\omega}=(i\omega+\Delta\tau_y\sigma_y)(1-1/\Gamma_{k,\omega})-\delta\mu_{k,\omega}\tau_z,
\end{equation}
where we define
\begin{gather}
\Gamma_{k,\omega}=\biggl(1+\frac{\gamma}{\Omega\varphi[\cosh(2\chi d)-\cos(2\zeta d)]}\biggl\{\sinh(2\chi d)-\cos(2\zeta z_{2D})\sinh[2\chi(d-z_{2D})]-\cos[2\zeta(d-z_{2D})]\sinh(2\chi z_{2D})\biggr\}\biggr)^{-1}, \nonumber \\
\delta\mu_{k,\omega}=-\frac{\gamma}{\varphi[\cosh(2\chi d)-\cos(2\zeta d)]}\biggl\{\sin(2\zeta d)-\sin(2\zeta z_{2D})\cosh[2\chi(d-z_{2D})]-\sin[2\zeta(d-z_{2D})]\cosh(2\chi z_{2D})\biggr\}, \label{effparams}
\end{gather}
\end{widetext}
with $\gamma=t^2/v_F$ an energy scale determined by the tunneling strength. The quantity $\Gamma_{k,\omega}$ can be interpreted as an effective quasiparticle weight, as it takes values $0<\Gamma<1$, and is responsible for inducing superconductivity in the 2DEG, while $\delta\mu_{k,\omega}$ corresponds to a tunneling-induced shift in the effective chemical potential of the 2DEG. Quite surprisingly, the self-energy Eqs.~\eqref{selfenergy2}--\eqref{effparams} coincides with that of a nanowire coupled to a two-dimensional superconductor with finite width as found in Ref.~\cite{Reeg:2017_3}, with the simple replacement of a 1D momentum by the magnitude of a 2D momentum.

\section{Induced Gap and band shift} \label{Sec3}

Using the self-energy derived in Sec.~\ref{Sec2}, we first calculate the size of the proximity-induced gap in the 2DEG. Once we find an expression for the gap, we estimate the tunneling strength needed in order for the gap in the 2DEG to be comparable to that in the superconductor. We then add a Zeeman term to the Hamiltonian of the 2DEG and estimate the Zeeman energy needed to reach the topological phase in such a setup.

It is convenient to work in the chiral basis in which the normal Green's function of the 2DEG is diagonal. To this end, we introduce a unitary transformation
\begin{equation}
U_\kk=\frac{1}{\sqrt{2}}\begin{pmatrix}
	1 & 1 & 0 & 0 \\
	-ie^{i\phi_\kk} & ie^{i\phi_\kk} & 0 & 0 \\
	0 & 0 & 1 & 1 \\
	0 & 0 & -ie^{-i\phi_\kk} & ie^{-i\phi_\kk}
	\end{pmatrix},
\end{equation}
with $\phi_\kk=\tan^{-1}(k_y/k_x)$, which can be used to convert between the spin ($\sigma$) and chiral ($\lambda$) bases, $G_{\kk,\omega}^\lambda=U^\dagger_\kk G_{\kk,\omega}^\sigma U_\kk$. The Green's function in the spin basis is given by $G_{\kk,\omega}^\sigma=(i\omega-\mathcal{H}_\kk^{2D}-\Sigma_{k,\omega})^{-1}$. Rotating to the chiral basis, we find a Green's function given by
\begin{equation} \label{Gchiral}
G_{\kk,\omega}^\lambda=\begin{pmatrix}
	\frac{-i\tilde\omega-\tilde\xi_+}{\tilde\omega^2+\tilde\xi_+^2+\tilde\Delta^2} & 0 & \frac{i\tilde\Delta e^{-i\phi_\kk}}{\tilde\omega^2+\tilde\xi_+^2+\tilde\Delta^2} & 0 \\
	0 & \frac{-i\tilde\omega-\tilde\xi_-}{\tilde\omega^2+\tilde\xi_-^2+\tilde\Delta^2} & 0 & \frac{-i\tilde\Delta e^{-i\phi_\kk}}{\tilde\omega^2+\tilde\xi_-^2+\tilde\Delta^2} \\
	\frac{-i\tilde\Delta e^{i\phi_\kk}}{\tilde\omega^2+\tilde\xi_+^2+\tilde\Delta^2} & 0 & \frac{-i\tilde\omega+\tilde\xi_+}{\tilde\omega^2+\tilde\xi_+^2+\tilde\Delta^2} & 0 \\
	0 & \frac{i\tilde\Delta e^{i\phi_\kk}}{\tilde\omega^2+\tilde\xi_-^2+\tilde\Delta^2} & 0 & \frac{-i\tilde\omega+\tilde\xi_-}{\tilde\omega^2+\tilde\xi_-^2+\tilde\Delta^2}
	\end{pmatrix},
\end{equation}
where $\tilde\omega=\omega/\Gamma_{k,\omega}$, $\tilde\xi_\pm=\xi_k-\delta\mu_{k,\omega}\pm\alpha k$, and $\tilde\Delta=\Delta(1/\Gamma_{k,\omega}-1)$. The spin-singlet pairing induced by the superconductor appears as intraband chiral $p$-wave pairing (of the form $p_x\pm ip_y$) when expressed in the chiral basis.

Before continuing, let us simplify the parameters $\Gamma_{k,\omega}$ and $\delta\mu_{k,\omega}$. We will focus on the limit where the thickness of the superconducting layer is much smaller than its coherence length, $d\ll \xi_s$ (equivalently, $\Delta\ll\delta E_s$), and where the normal layer is located close to the edge of the superconductor, $k_Fz_{2D}\ll 1$. Due to the large Fermi surface mismatch between the 2DEG and superconductor, we must have $k\ll k_F$ (or, equivalently, $\varphi\approx1$); in the following, we neglect the momentum dependence by setting $\varphi=1$ (which is justified as long as we only consider momenta $k\ll1/\sqrt{k_Fd}$). In the limit $\omega\ll\delta E_s$, the parameters simplify to
\begin{equation} \label{effparams2}
\begin{aligned}
\Gamma&=\left(1+\frac{2\pi\gamma(k_Fz_{2D})^2}{\delta E_s\sin^2(k_Fd)}\right)^{-1}, \\
\delta\mu&=2\gamma(k_Fz_{2D})[1-(k_Fz_{2D})\cot(k_Fd)],
\end{aligned}
\end{equation}
where we drop the subscript $(k,\omega)$ because both $\Gamma$ and $\delta\mu$ are now independent of frequency and momentum. In expanding Eq.~\eqref{effparams} to arrive at Eq.~\eqref{effparams2}, we assumed that $|\sin(k_Fd)|\gg\Delta/\delta E_s$ (therefore, these expressions break down when $k_Fd/\pi\to n$, with $n\in\mathbb{Z}^+$).


The spectrum of the proximitized 2DEG is determined by the poles of the retarded Green's function. After analytic continuation $i\omega\to E+i0^+$, we find two branches of the spectrum from Eq.~\eqref{Gchiral} given by
\begin{equation} \label{spectrum}
E_\pm^2(k)=\Gamma^2\left(\frac{k^2}{2m_{2D}}-\mu_\text{eff}\pm\alpha k\right)^2+\Delta^2(1-\Gamma)^2,
\end{equation}
where $\mu_\text{eff}=\mu_{2D}+\delta\mu$ is an effective chemical potential of the 2DEG. The spectrum describes an $s$-wave superconductor with Rashba-split bands and an excitation gap
\begin{equation} \label{gapeq}
E_g=\Delta(1-\Gamma).
\end{equation}
We see that the size of the excitation gap is determined by the parameter $\Gamma$. When $\Gamma\ll1$, the full bulk gap of the superconductor is induced in the 2DEG, while for $(1-\Gamma)\ll1$, a very small gap is induced. In order to have an induced gap comparable (but not equal) to the bulk gap, we require $\Gamma\sim(1-\Gamma)\sim1$ [\emph{i.e.}, neither $\Gamma\ll1$ nor $(1-\Gamma)\ll1$ is satisfied]. However, to realize this situation requires a tunneling strength
\begin{equation} \label{gammaestimate}
\gamma\sim \delta E_s,
\end{equation}
where we have assumed that $(k_Fz_{2D})^2/\sin^2(k_Fd)\sim1$. If the tunneling strength is chosen as in Eq.~\eqref{gammaestimate}, the band shift measured at $k=0$, $E_\pm(0)$, is
\begin{equation} \label{deltamuestimate}
E_\pm(0)\sim\Gamma\delta\mu\sim\delta E_s.
\end{equation}
Therefore, the scale of the band shift is also set by the level spacing in the thin superconducting layer. We note that while the quantity $\delta\mu$ is bounded only by the chemical potential of the superconductor $\mu_s$ (as the tunneling Hamiltonian approach itself should break down for $\gamma\sim\mu_s$), the band shift saturates to $E_\pm(0)\sim\delta E_s$ in the limit $\gamma\gg\delta E_s$ (where $\Gamma\ll1$).

\begin{figure}[t!]
\centering
\includegraphics[width=\linewidth]{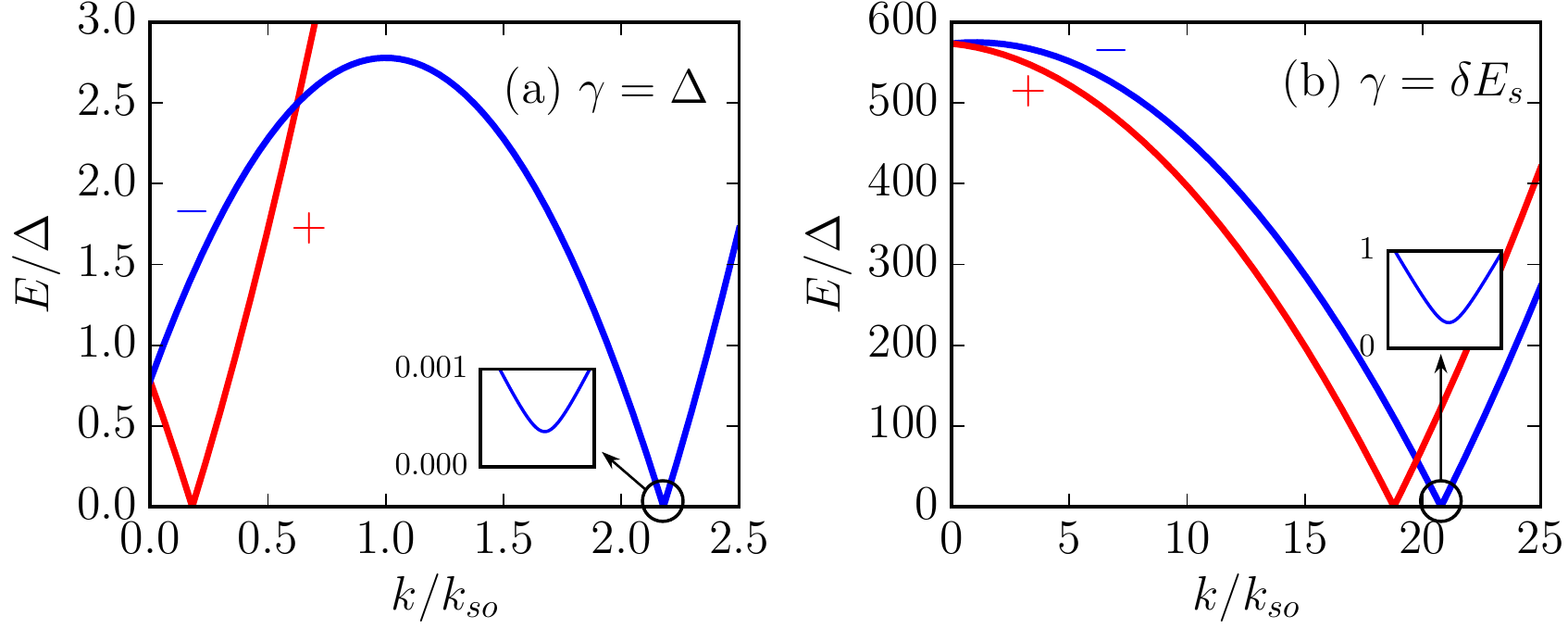}
\caption{\label{spectrumplot} Spectrum of a 2DEG coupled to a thin superconducting layer [see Eq.~\eqref{spectrum}] for (a) $\gamma=\Delta$ (corresponding to $\Gamma=0.9996$ and $\delta\mu=0.78\Delta$) and (b) $\gamma=\delta E_s$ (corresponding to $\Gamma=0.735$ and $\delta\mu=780\Delta$). When tunneling is weak [as in (a)], the band shift is rather small but the induced gap is negligible. If tunneling is strong enough to open a sizable gap [as in (b)], the band shift is very large (note that the band shift is given by $E_\pm(0)\sim\Gamma\delta\mu$ rather than $\delta\mu$). In both plots, $E_{so}=2\Delta$, $\delta E_s=1000\Delta$, $\mu_{2D}=0$, $k_Fd/\pi=48.75$, and $k_Fz_{2D}=0.3$. Here $k_{so}=m\alpha$ is the spin-orbit momentum. Note that although, in the insets, we show only the induced gap on the ``$-$"-subband, there is an equally large gap induced on the ``+"-subband.}
\end{figure}

We plot the spectrum of the 2DEG [see Eq.~\eqref{spectrum}] in Fig.~\ref{spectrumplot}. In the weak-coupling limit [Fig.~\ref{spectrumplot}(a)], there is a rather small band shift but a negligible superconducting gap is opened in the 2DEG. In the strong-coupling limit [Fig.~\ref{spectrumplot}(b)], we show that while a larger gap is induced, the band shift is very large.

\section{Topological Transition}
We now add a Zeeman splitting $\Delta_Z$ to the Hamiltonian of the 2DEG such that
\begin{equation}
\mathcal{H}_\kk^{2D}=\xi_k\tau_z+\alpha(k_y\sigma_x-k_x\tau_z\sigma_y)-\Delta_Z\tau_z\sigma_z.
\end{equation}
Such a Zeeman splitting can arise due to the application of an out-of-plane magnetic field \cite{Sato:2009,Sato:2009b} (though orbital effects are not incorporated here) or due to proximity with a magnetic insulator \cite{Sau:2010}. Also, it is possible to apply an in-plane magnetic field (to avoid unwanted orbital effects) to reach the topological phase if the 2DEG has a finite Dresselhaus SOI, as shown in Ref.~\cite{Alicea:2010}. An in-plane magnetic field in the presence of only Rashba SOI is not sufficient to reach the topological phase because it does not open a gap in the Rashba spectrum. The spectrum in the presence of the Zeeman splitting, which again is determined by poles in the retarded Green's function $G^R_{\kk,E}=(E-\mathcal{H}_\kk^{2D}-\Sigma+i0^+)^{-1}$, is given by
\begin{equation}
\begin{aligned}
E^2&=\Gamma^2[\Delta_Z^2+(\xi_k-\delta\mu)^2+\alpha^2k^2]+E_g^2 \\
	&\pm2\Gamma\sqrt{\Delta_Z^2E_g^2+\Gamma^2(\xi_k-\delta\mu)^2(\Delta_Z^2+\alpha^2k^2)},
\end{aligned}
\end{equation}
where we have used $E_g=\Delta(1-\Gamma)$ as in Eq.~\eqref{gapeq}. Therefore, we find a gap-closing topological transition at $k=0$ for the critical Zeeman splitting
\begin{equation} \label{toptransition}
\Gamma\Delta_Z^c=\sqrt{\Gamma^2(\mu_{2D}+\delta\mu)^2+E_g^2}.
\end{equation}
In the case of a very large band shift, $\Gamma\delta\mu\gg E_g$ and $\delta\mu\gg\mu_{2D}$, the topological transition is given by $\Delta_Z^c=\delta\mu$ [note that $\Gamma$ drops out of Eq.~\eqref{toptransition} in this limit].

We now provide an estimate of the Zeeman splitting at which we expect the $k=0$ gap-closing transition to occur experimentally in an Al/InAs 2DEG heterostructure. Given the thickness of the superconducting Al layer of $d=10$ nm \cite{Shabani:2016}, we estimate a level spacing of $\delta E_s=\pi\hbar v_F/d=413$ meV (taking $v_F=2\times10^6$ m/s). Therefore, if a sizable gap is induced in the 2DEG, as observed experimentally, typical values for the band shift are of the same order of magnitude as the level spacing, $\Gamma\delta\mu\sim400$ meV. Then, provided that the chemical potential cannot be controlled over such a large scale by external gates, the critical Zeeman splitting needed to reach the topological phase is $\Delta_Z^c=\delta\mu\sim400$ meV. Such a large Zeeman splitting cannot be achieved in the 2DEG without destroying superconductivity in the thin layer. We also note the possibility that, by coincidence, the band shift vanishes (or becomes small); from {Eq.~\eqref{effparams2}}, we see that $\delta\mu=0$ if $k_Fd=\cot^{-1}(1/k_Fz_{2D})+n\pi$ (for $n\in\mathbb{Z}$). In this special case, which requires the thickness of the superconducting layer to be finely tuned on the scale of its Fermi wavelength, there is no band shift to prevent one from tuning into a topological phase. However, for most devices, the large band shift makes it very challenging to realize a topological phase.

\section{Increasing thickness of superconducting layer}

\begin{figure}[t!]
\centering
\includegraphics[width=\linewidth]{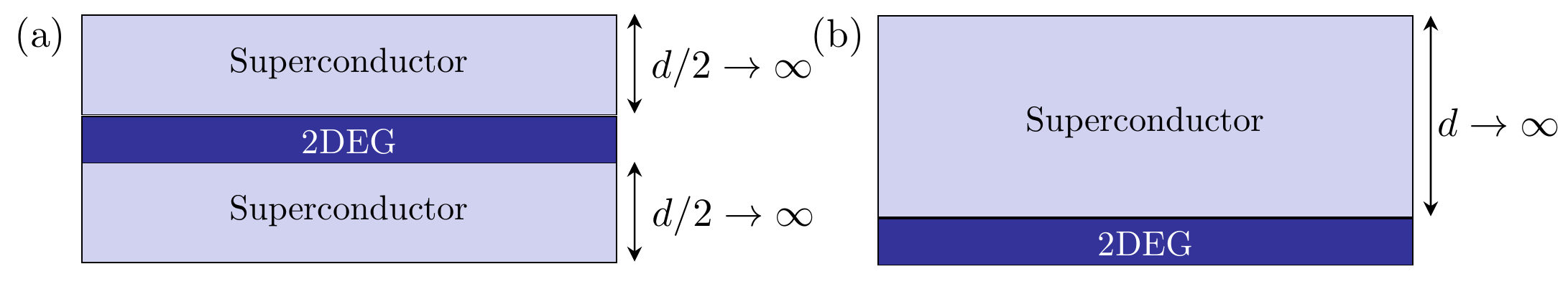}
\caption{\label{bulkcase} (a) Evaluating the self-energy with the Green's function of a bulk superconductor [see Eq.~\eqref{effparamsbulk}] corresponds to a 2DEG embedded within an infinitely large superconductor. (b) Evaluating the self-energy with the Green's function of a semi-infinite superconductor [see Eq.~\eqref{effparamssi}] corresponds to a 2DEG placed on the surface of an infinitely large superconductor.}
\end{figure}

The self-energy appearing most frequently in the literature to describe proximitized nanowires and 2DEGs \cite{Sau:2010prox,Potter:2011,Kopnin:2011,Zyuzin:2013,vanHeck:2016,Reeg:2017_2}, and which has often been used in interpreting experimental results \cite{Deng:2016,Zhang:2017_2}, is that induced by a bulk superconductor,
\begin{equation} \label{effparamsbulk}
\begin{aligned}
\Gamma_{\text{bulk}}&=\left(1+\gamma/\Omega\right)^{-1}, \\
\delta\mu_{\text{bulk}}&=0.
\end{aligned}
\end{equation}
Equation \eqref{effparamsbulk} can be obtained by setting $z_{2D}=d/2$ and taking the limit $d\to\infty$ in Eq.~\eqref{effparams} [or, as is usually done, by substituting the bulk Green's function in Eq.~\eqref{bulkG} when evaluating the self-energy in Eq.~\eqref{selfenergy}]. Hence, this self-energy describes a 2DEG embedded \emph{within} a bulk superconductor, as shown in Fig.~\ref{bulkcase}(a). To describe the case where a 2DEG is placed at the \emph{surface} of a bulk superconductor [as shown in Fig.~\ref{bulkcase}(b)], the limit $d\to\infty$ should be taken in Eq.~\eqref{effparams} while keeping $z_{2D}$ finite [or, equivalently, substituting the Green's function of a semi-infinite (SI) superconductor when evaluating the self-energy in Eq.~\eqref{selfenergy}]. For this case, we obtain
\begin{equation} \label{effparamssi}
\begin{aligned}
\Gamma_\text{SI}&=\left(1+\frac{\gamma}{\Omega}\left\{1-\cos(2\zeta z_{2D})e^{-2\chi z_{2D}}\right\}\right)^{-1}, \\
\delta\mu_\text{SI}&=\gamma\sin(2\zeta z_{2D})e^{-2\chi z_{2D}}.
\end{aligned}
\end{equation}
The most notable difference is the presence of a nonzero band shift in the semi-infinite case. However, this band shift is significantly reduced compared to the case of a thin superconducting layer, as it saturates to $E_\pm(0)\sim\Gamma_\text{SI}\delta\mu_\text{SI}\sim\Delta$ in the limit $\gamma\gg\Delta$.

While it may naively seem that a topological phase can be much more easily realized by simply increasing the thickness of the superconducting layer in order to reduce the band shift induced on the 2DEG, this is not the case. Crucially, both the the bulk and semi-infinite self-energies give the ratio $\gamma/\Delta$ as the relevant parameter determining whether the system is in the weak-coupling [$(1-\Gamma)\ll1$, or equivalently $E_g\ll\Delta$] or strong-coupling [$(1-\Gamma)\sim1$, or equivalently $E_g\sim\Delta$] limit. This is in stark contrast to the limit of a thin superconducting layer, where a tunneling energy $\gamma\sim\delta E_s\gg\Delta$ is required to open a gap $E_g\sim\Delta$ in the 2DEG. Therefore, because the tunneling energy $\gamma$ is a property of the interface and should not be expected to change as the thickness of the superconducting layer is increased, this energy is fixed to $\gamma\sim\delta E_s$ provided that the interface is transparent enough to induce a gap in the thin-layer limit (as seen in the experiments). If the thickness of the superconductor is increased, such that $d\gg\xi_s$, the system will be deep within the strong-coupling limit; from Eqs.~\eqref{effparamsbulk} and \eqref{effparamssi}, we find $\Gamma\sim\Delta/\gamma\ll1$. The critical Zeeman splitting needed to induce a topological phase [see Eq.~\eqref{toptransition}] therefore is given by $\Delta_Z^c\sim\Delta/\Gamma\sim\gamma\sim400$ meV. We note that in the case of a thin superconducting layer, the topological transition is pushed to large Zeeman splitting by very large $\delta\mu$, which could possibly be compensated for if the chemical potential $\mu_{2D}$ has a large range of tunability; in the case of a bulk system, the topological transition is pushed to large Zeeman splitting by very small $\Gamma$, which cannot be affected by tuning $\mu_{2D}$. Hence, even if the thickness $d$ of the superconducting layer is made infinite, the topological phase transition is determined by the interfacial tunneling energy. In order to more reliably induce a topological phase, a much weaker coupling between a 2DEG and a bulk superconductor (such that $\gamma\lesssim\Delta$) should be sought. We note that this result applies to the 1D model considered in Ref.~\cite{Reeg:2017_3} as well.

\section{Conclusions} \label{Sec6}
We have studied the proximity effect in a two-dimensional electron gas (2DEG) strongly coupled to a thin superconducting layer, showing that the detrimental band shift shown in Refs.~\cite{Reeg:2017_3,Reeg:2018} to dominate the proximity effect in wires is also crucial in 2DEGs. In order to induce a sizable gap in the 2DEG, the tunneling energy scale must overcome the large level spacing within the superconductor. However, introducing such a large energy scale to the semiconductor induces a large band shift that makes it challenging to realize a topological phase. This challenge cannot be alleviated by simply increasing the thickness of the superconducting layer but requires a significant weakening of the proximity coupling afforded by the epitaxial interface.

\acknowledgments
This work was supported by the Swiss National Science Foundation and the NCCR QSIT.

\bibliography{bibStrongCoupling2D}

\begin{thebibliography}{60}%
\makeatletter
\providecommand \@ifxundefined [1]{%
 \@ifx{#1\undefined}
}%
\providecommand \@ifnum [1]{%
 \ifnum #1\expandafter \@firstoftwo
 \else \expandafter \@secondoftwo
 \fi
}%
\providecommand \@ifx [1]{%
 \ifx #1\expandafter \@firstoftwo
 \else \expandafter \@secondoftwo
 \fi
}%
\providecommand \natexlab [1]{#1}%
\providecommand \enquote  [1]{``#1''}%
\providecommand \bibnamefont  [1]{#1}%
\providecommand \bibfnamefont [1]{#1}%
\providecommand \citenamefont [1]{#1}%
\providecommand \href@noop [0]{\@secondoftwo}%
\providecommand \href [0]{\begingroup \@sanitize@url \@href}%
\providecommand \@href[1]{\@@startlink{#1}\@@href}%
\providecommand \@@href[1]{\endgroup#1\@@endlink}%
\providecommand \@sanitize@url [0]{\catcode `\\12\catcode `\$12\catcode
  `\&12\catcode `\#12\catcode `\^12\catcode `\_12\catcode `\%12\relax}%
\providecommand \@@startlink[1]{}%
\providecommand \@@endlink[0]{}%
\providecommand \url  [0]{\begingroup\@sanitize@url \@url }%
\providecommand \@url [1]{\endgroup\@href {#1}{\urlprefix }}%
\providecommand \urlprefix  [0]{URL }%
\providecommand \Eprint [0]{\href }%
\providecommand \doibase [0]{http://dx.doi.org/}%
\providecommand \selectlanguage [0]{\@gobble}%
\providecommand \bibinfo  [0]{\@secondoftwo}%
\providecommand \bibfield  [0]{\@secondoftwo}%
\providecommand \translation [1]{[#1]}%
\providecommand \BibitemOpen [0]{}%
\providecommand \bibitemStop [0]{}%
\providecommand \bibitemNoStop [0]{.\EOS\space}%
\providecommand \EOS [0]{\spacefactor3000\relax}%
\providecommand \BibitemShut  [1]{\csname bibitem#1\endcsname}%
\let\auto@bib@innerbib\@empty
\bibitem [{\citenamefont {Kitaev}(2001)}]{Kitaev:2001}%
  \BibitemOpen
  \bibfield  {author} {\bibinfo {author} {\bibfnamefont {A.~Y.}\ \bibnamefont
  {Kitaev}},\ }\href {http://stacks.iop.org/1063-7869/44/i=10S/a=S29}
  {\bibfield  {journal} {\bibinfo  {journal} {Physics-Uspekhi}\ }\textbf
  {\bibinfo {volume} {{\bf 44}}},\ \bibinfo {pages} {131} (\bibinfo {year}
  {2001})}\BibitemShut {NoStop}%
\bibitem [{\citenamefont {Nayak}\ \emph {et~al.}(2008)\citenamefont {Nayak},
  \citenamefont {Simon}, \citenamefont {Stern}, \citenamefont {Freedman},\ and\
  \citenamefont {Das~Sarma}}]{Nayak:2008}%
  \BibitemOpen
  \bibfield  {author} {\bibinfo {author} {\bibfnamefont {C.}~\bibnamefont
  {Nayak}}, \bibinfo {author} {\bibfnamefont {S.~H.}\ \bibnamefont {Simon}},
  \bibinfo {author} {\bibfnamefont {A.}~\bibnamefont {Stern}}, \bibinfo
  {author} {\bibfnamefont {M.}~\bibnamefont {Freedman}}, \ and\ \bibinfo
  {author} {\bibfnamefont {S.}~\bibnamefont {Das~Sarma}},\ }\href
  {http://link.aps.org/doi/10.1103/RevModPhys.80.1083} {\bibfield  {journal}
  {\bibinfo  {journal} {Rev. Mod. Phys.}\ }\textbf {\bibinfo {volume} {{\bf
  80}}},\ \bibinfo {pages} {1083} (\bibinfo {year} {2008})}\BibitemShut
  {NoStop}%
\bibitem [{\citenamefont {Alicea}(2012)}]{Alicea:2012}%
  \BibitemOpen
  \bibfield  {author} {\bibinfo {author} {\bibfnamefont {J.}~\bibnamefont
  {Alicea}},\ }\href {http://stacks.iop.org/0034-4885/75/i=7/a=076501}
  {\bibfield  {journal} {\bibinfo  {journal} {Rep. Prog. Phys.}\ }\textbf
  {\bibinfo {volume} {{\bf 75}}},\ \bibinfo {pages} {076501} (\bibinfo {year}
  {2012})}\BibitemShut {NoStop}%
\bibitem [{\citenamefont {Sato}\ and\ \citenamefont
  {Fujimoto}(2009)}]{Sato:2009}%
  \BibitemOpen
  \bibfield  {author} {\bibinfo {author} {\bibfnamefont {M.}~\bibnamefont
  {Sato}}\ and\ \bibinfo {author} {\bibfnamefont {S.}~\bibnamefont
  {Fujimoto}},\ }\href {https://link.aps.org/doi/10.1103/PhysRevB.79.094504}
  {\bibfield  {journal} {\bibinfo  {journal} {Phys. Rev. B}\ }\textbf {\bibinfo
  {volume} {{\bf 79}}},\ \bibinfo {pages} {094504} (\bibinfo {year}
  {2009})}\BibitemShut {NoStop}%
\bibitem [{\citenamefont {Sato}\ \emph {et~al.}(2009)\citenamefont {Sato},
  \citenamefont {Takahashi},\ and\ \citenamefont {Fujimoto}}]{Sato:2009b}%
  \BibitemOpen
  \bibfield  {author} {\bibinfo {author} {\bibfnamefont {M.}~\bibnamefont
  {Sato}}, \bibinfo {author} {\bibfnamefont {Y.}~\bibnamefont {Takahashi}}, \
  and\ \bibinfo {author} {\bibfnamefont {S.}~\bibnamefont {Fujimoto}},\ }\href
  {https://link.aps.org/doi/10.1103/PhysRevLett.103.020401} {\bibfield
  {journal} {\bibinfo  {journal} {Phys. Rev. Lett.}\ }\textbf {\bibinfo
  {volume} {{\bf 103}}},\ \bibinfo {pages} {020401} (\bibinfo {year}
  {2009})}\BibitemShut {NoStop}%
\bibitem [{\citenamefont {Lutchyn}\ \emph {et~al.}(2010)\citenamefont
  {Lutchyn}, \citenamefont {Sau},\ and\ \citenamefont
  {Das~Sarma}}]{Lutchyn:2010}%
  \BibitemOpen
  \bibfield  {author} {\bibinfo {author} {\bibfnamefont {R.~M.}\ \bibnamefont
  {Lutchyn}}, \bibinfo {author} {\bibfnamefont {J.~D.}\ \bibnamefont {Sau}}, \
  and\ \bibinfo {author} {\bibfnamefont {S.}~\bibnamefont {Das~Sarma}},\ }\href
  {\doibase 10.1103/PhysRevLett.105.077001} {\bibfield  {journal} {\bibinfo
  {journal} {Phys. Rev. Lett.}\ }\textbf {\bibinfo {volume} {{\bf 105}}},\
  \bibinfo {pages} {077001} (\bibinfo {year} {2010})}\BibitemShut {NoStop}%
\bibitem [{\citenamefont {Oreg}\ \emph {et~al.}(2010)\citenamefont {Oreg},
  \citenamefont {Refael},\ and\ \citenamefont {von Oppen}}]{Oreg:2010}%
  \BibitemOpen
  \bibfield  {author} {\bibinfo {author} {\bibfnamefont {Y.}~\bibnamefont
  {Oreg}}, \bibinfo {author} {\bibfnamefont {G.}~\bibnamefont {Refael}}, \ and\
  \bibinfo {author} {\bibfnamefont {F.}~\bibnamefont {von Oppen}},\ }\href
  {\doibase 10.1103/PhysRevLett.105.177002} {\bibfield  {journal} {\bibinfo
  {journal} {Phys. Rev. Lett.}\ }\textbf {\bibinfo {volume} {{\bf 105}}},\
  \bibinfo {pages} {177002} (\bibinfo {year} {2010})}\BibitemShut {NoStop}%
\bibitem [{\citenamefont {Sau}\ \emph {et~al.}(2010{\natexlab{a}})\citenamefont
  {Sau}, \citenamefont {Lutchyn}, \citenamefont {Tewari},\ and\ \citenamefont
  {Das~Sarma}}]{Sau:2010}%
  \BibitemOpen
  \bibfield  {author} {\bibinfo {author} {\bibfnamefont {J.~D.}\ \bibnamefont
  {Sau}}, \bibinfo {author} {\bibfnamefont {R.~M.}\ \bibnamefont {Lutchyn}},
  \bibinfo {author} {\bibfnamefont {S.}~\bibnamefont {Tewari}}, \ and\ \bibinfo
  {author} {\bibfnamefont {S.}~\bibnamefont {Das~Sarma}},\ }\href
  {http://link.aps.org/doi/10.1103/PhysRevLett.104.040502} {\bibfield
  {journal} {\bibinfo  {journal} {Phys. Rev. Lett.}\ }\textbf {\bibinfo
  {volume} {{\bf 104}}},\ \bibinfo {pages} {040502} (\bibinfo {year}
  {2010}{\natexlab{a}})}\BibitemShut {NoStop}%
\bibitem [{\citenamefont {Alicea}(2010)}]{Alicea:2010}%
  \BibitemOpen
  \bibfield  {author} {\bibinfo {author} {\bibfnamefont {J.}~\bibnamefont
  {Alicea}},\ }\href {http://link.aps.org/doi/10.1103/PhysRevB.81.125318}
  {\bibfield  {journal} {\bibinfo  {journal} {Phys. Rev. B}\ }\textbf {\bibinfo
  {volume} {{\bf 81}}},\ \bibinfo {pages} {125318} (\bibinfo {year}
  {2010})}\BibitemShut {NoStop}%
\bibitem [{\citenamefont {Chevallier}\ \emph {et~al.}(2012)\citenamefont
  {Chevallier}, \citenamefont {Sticlet}, \citenamefont {Simon},\ and\
  \citenamefont {Bena}}]{Chevallier:2012}%
  \BibitemOpen
  \bibfield  {author} {\bibinfo {author} {\bibfnamefont {D.}~\bibnamefont
  {Chevallier}}, \bibinfo {author} {\bibfnamefont {D.}~\bibnamefont {Sticlet}},
  \bibinfo {author} {\bibfnamefont {P.}~\bibnamefont {Simon}}, \ and\ \bibinfo
  {author} {\bibfnamefont {C.}~\bibnamefont {Bena}},\ }\href
  {https://link.aps.org/doi/10.1103/PhysRevB.85.235307} {\bibfield  {journal}
  {\bibinfo  {journal} {Phys. Rev. B}\ }\textbf {\bibinfo {volume} {{\bf
  85}}},\ \bibinfo {pages} {235307} (\bibinfo {year} {2012})}\BibitemShut
  {NoStop}%
\bibitem [{\citenamefont {Halperin}\ \emph {et~al.}(2012)\citenamefont
  {Halperin}, \citenamefont {Oreg}, \citenamefont {Stern}, \citenamefont
  {Refael}, \citenamefont {Alicea},\ and\ \citenamefont {von
  Oppen}}]{Halperin:2012}%
  \BibitemOpen
  \bibfield  {author} {\bibinfo {author} {\bibfnamefont {B.~I.}\ \bibnamefont
  {Halperin}}, \bibinfo {author} {\bibfnamefont {Y.}~\bibnamefont {Oreg}},
  \bibinfo {author} {\bibfnamefont {A.}~\bibnamefont {Stern}}, \bibinfo
  {author} {\bibfnamefont {G.}~\bibnamefont {Refael}}, \bibinfo {author}
  {\bibfnamefont {J.}~\bibnamefont {Alicea}}, \ and\ \bibinfo {author}
  {\bibfnamefont {F.}~\bibnamefont {von Oppen}},\ }\href
  {https://link.aps.org/doi/10.1103/PhysRevB.85.144501} {\bibfield  {journal}
  {\bibinfo  {journal} {Phys. Rev. B}\ }\textbf {\bibinfo {volume} {{\bf
  85}}},\ \bibinfo {pages} {144501} (\bibinfo {year} {2012})}\BibitemShut
  {NoStop}%
\bibitem [{\citenamefont {Sticlet}\ \emph {et~al.}(2012)\citenamefont
  {Sticlet}, \citenamefont {Bena},\ and\ \citenamefont {Simon}}]{Sticlet:2012}%
  \BibitemOpen
  \bibfield  {author} {\bibinfo {author} {\bibfnamefont {D.}~\bibnamefont
  {Sticlet}}, \bibinfo {author} {\bibfnamefont {C.}~\bibnamefont {Bena}}, \
  and\ \bibinfo {author} {\bibfnamefont {P.}~\bibnamefont {Simon}},\ }\href
  {https://link.aps.org/doi/10.1103/PhysRevLett.108.096802} {\bibfield
  {journal} {\bibinfo  {journal} {Phys. Rev. Lett.}\ }\textbf {\bibinfo
  {volume} {{\bf 108}}},\ \bibinfo {pages} {096802} (\bibinfo {year}
  {2012})}\BibitemShut {NoStop}%
\bibitem [{\citenamefont {Klinovaja}\ \emph
  {et~al.}(2012{\natexlab{a}})\citenamefont {Klinovaja}, \citenamefont
  {Stano},\ and\ \citenamefont {Loss}}]{Klinovaja:2012}%
  \BibitemOpen
  \bibfield  {author} {\bibinfo {author} {\bibfnamefont {J.}~\bibnamefont
  {Klinovaja}}, \bibinfo {author} {\bibfnamefont {P.}~\bibnamefont {Stano}}, \
  and\ \bibinfo {author} {\bibfnamefont {D.}~\bibnamefont {Loss}},\ }\href
  {http://link.aps.org/doi/10.1103/PhysRevLett.109.236801} {\bibfield
  {journal} {\bibinfo  {journal} {Phys. Rev. Lett.}\ }\textbf {\bibinfo
  {volume} {{\bf 109}}},\ \bibinfo {pages} {236801} (\bibinfo {year}
  {2012}{\natexlab{a}})}\BibitemShut {NoStop}%
\bibitem [{\citenamefont {Klinovaja}\ \emph
  {et~al.}(2012{\natexlab{b}})\citenamefont {Klinovaja}, \citenamefont
  {Gangadharaiah},\ and\ \citenamefont {Loss}}]{Klinovaja:2012c}%
  \BibitemOpen
  \bibfield  {author} {\bibinfo {author} {\bibfnamefont {J.}~\bibnamefont
  {Klinovaja}}, \bibinfo {author} {\bibfnamefont {S.}~\bibnamefont
  {Gangadharaiah}}, \ and\ \bibinfo {author} {\bibfnamefont {D.}~\bibnamefont
  {Loss}},\ }\href {https://link.aps.org/doi/10.1103/PhysRevLett.108.196804}
  {\bibfield  {journal} {\bibinfo  {journal} {Phys. Rev. Lett.}\ }\textbf
  {\bibinfo {volume} {{\bf 108}}},\ \bibinfo {pages} {196804} (\bibinfo {year}
  {2012}{\natexlab{b}})}\BibitemShut {NoStop}%
\bibitem [{\citenamefont {Prada}\ \emph {et~al.}(2012)\citenamefont {Prada},
  \citenamefont {San-Jose},\ and\ \citenamefont {Aguado}}]{Prada:2012}%
  \BibitemOpen
  \bibfield  {author} {\bibinfo {author} {\bibfnamefont {E.}~\bibnamefont
  {Prada}}, \bibinfo {author} {\bibfnamefont {P.}~\bibnamefont {San-Jose}}, \
  and\ \bibinfo {author} {\bibfnamefont {R.}~\bibnamefont {Aguado}},\ }\href
  {https://link.aps.org/doi/10.1103/PhysRevB.86.180503} {\bibfield  {journal}
  {\bibinfo  {journal} {Phys. Rev. B}\ }\textbf {\bibinfo {volume} {{\bf
  86}}},\ \bibinfo {pages} {180503} (\bibinfo {year} {2012})}\BibitemShut
  {NoStop}%
\bibitem [{\citenamefont {Dom\'{\i}nguez}\ \emph {et~al.}(2012)\citenamefont
  {Dom\'{\i}nguez}, \citenamefont {Hassler},\ and\ \citenamefont
  {Platero}}]{Dominguez:2012}%
  \BibitemOpen
  \bibfield  {author} {\bibinfo {author} {\bibfnamefont {F.}~\bibnamefont
  {Dom\'{\i}nguez}}, \bibinfo {author} {\bibfnamefont {F.}~\bibnamefont
  {Hassler}}, \ and\ \bibinfo {author} {\bibfnamefont {G.}~\bibnamefont
  {Platero}},\ }\href {https://link.aps.org/doi/10.1103/PhysRevB.86.140503}
  {\bibfield  {journal} {\bibinfo  {journal} {Phys. Rev. B}\ }\textbf {\bibinfo
  {volume} {{\bf 86}}},\ \bibinfo {pages} {140503} (\bibinfo {year}
  {2012})}\BibitemShut {NoStop}%
\bibitem [{\citenamefont {Klinovaja}\ and\ \citenamefont
  {Loss}(2013)}]{Klinovaja:2013b}%
  \BibitemOpen
  \bibfield  {author} {\bibinfo {author} {\bibfnamefont {J.}~\bibnamefont
  {Klinovaja}}\ and\ \bibinfo {author} {\bibfnamefont {D.}~\bibnamefont
  {Loss}},\ }\href {https://link.aps.org/doi/10.1103/PhysRevX.3.011008}
  {\bibfield  {journal} {\bibinfo  {journal} {Phys. Rev. X}\ }\textbf {\bibinfo
  {volume} {{\bf 3}}},\ \bibinfo {pages} {011008} (\bibinfo {year}
  {2013})}\BibitemShut {NoStop}%
\bibitem [{\citenamefont {DeGottardi}\ \emph {et~al.}(2013)\citenamefont
  {DeGottardi}, \citenamefont {Thakurathi}, \citenamefont {Vishveshwara},\ and\
  \citenamefont {Sen}}]{DeGottardi:2013}%
  \BibitemOpen
  \bibfield  {author} {\bibinfo {author} {\bibfnamefont {W.}~\bibnamefont
  {DeGottardi}}, \bibinfo {author} {\bibfnamefont {M.}~\bibnamefont
  {Thakurathi}}, \bibinfo {author} {\bibfnamefont {S.}~\bibnamefont
  {Vishveshwara}}, \ and\ \bibinfo {author} {\bibfnamefont {D.}~\bibnamefont
  {Sen}},\ }\href {https://link.aps.org/doi/10.1103/PhysRevB.88.165111}
  {\bibfield  {journal} {\bibinfo  {journal} {Phys. Rev. B}\ }\textbf {\bibinfo
  {volume} {{\bf 88}}},\ \bibinfo {pages} {165111} (\bibinfo {year}
  {2013})}\BibitemShut {NoStop}%
\bibitem [{\citenamefont {Maier}\ \emph {et~al.}(2014)\citenamefont {Maier},
  \citenamefont {Klinovaja},\ and\ \citenamefont {Loss}}]{Maier:2014}%
  \BibitemOpen
  \bibfield  {author} {\bibinfo {author} {\bibfnamefont {F.}~\bibnamefont
  {Maier}}, \bibinfo {author} {\bibfnamefont {J.}~\bibnamefont {Klinovaja}}, \
  and\ \bibinfo {author} {\bibfnamefont {D.}~\bibnamefont {Loss}},\ }\href
  {https://link.aps.org/doi/10.1103/PhysRevB.90.195421} {\bibfield  {journal}
  {\bibinfo  {journal} {Phys. Rev. B}\ }\textbf {\bibinfo {volume} {{\bf
  90}}},\ \bibinfo {pages} {195421} (\bibinfo {year} {2014})}\BibitemShut
  {NoStop}%
\bibitem [{\citenamefont {Vernek}\ \emph {et~al.}(2014)\citenamefont {Vernek},
  \citenamefont {Penteado}, \citenamefont {Seridonio},\ and\ \citenamefont
  {Egues}}]{Vernek:2014}%
  \BibitemOpen
  \bibfield  {author} {\bibinfo {author} {\bibfnamefont {E.}~\bibnamefont
  {Vernek}}, \bibinfo {author} {\bibfnamefont {P.~H.}\ \bibnamefont
  {Penteado}}, \bibinfo {author} {\bibfnamefont {A.~C.}\ \bibnamefont
  {Seridonio}}, \ and\ \bibinfo {author} {\bibfnamefont {J.~C.}\ \bibnamefont
  {Egues}},\ }\href {https://link.aps.org/doi/10.1103/PhysRevB.89.165314}
  {\bibfield  {journal} {\bibinfo  {journal} {Phys. Rev. B}\ }\textbf {\bibinfo
  {volume} {{\bf 89}}},\ \bibinfo {pages} {165314} (\bibinfo {year}
  {2014})}\BibitemShut {NoStop}%
\bibitem [{\citenamefont {Weithofer}\ \emph {et~al.}(2014)\citenamefont
  {Weithofer}, \citenamefont {Recher},\ and\ \citenamefont
  {Schmidt}}]{Weithofer:2014}%
  \BibitemOpen
  \bibfield  {author} {\bibinfo {author} {\bibfnamefont {L.}~\bibnamefont
  {Weithofer}}, \bibinfo {author} {\bibfnamefont {P.}~\bibnamefont {Recher}}, \
  and\ \bibinfo {author} {\bibfnamefont {T.~L.}\ \bibnamefont {Schmidt}},\
  }\href {https://link.aps.org/doi/10.1103/PhysRevB.90.205416} {\bibfield
  {journal} {\bibinfo  {journal} {Phys. Rev. B}\ }\textbf {\bibinfo {volume}
  {{\bf 90}}},\ \bibinfo {pages} {205416} (\bibinfo {year} {2014})}\BibitemShut
  {NoStop}%
\bibitem [{\citenamefont {Thakurathi}\ \emph {et~al.}(2015)\citenamefont
  {Thakurathi}, \citenamefont {Deb},\ and\ \citenamefont
  {Sen}}]{Thakurathi:2015}%
  \BibitemOpen
  \bibfield  {author} {\bibinfo {author} {\bibfnamefont {M.}~\bibnamefont
  {Thakurathi}}, \bibinfo {author} {\bibfnamefont {O.}~\bibnamefont {Deb}}, \
  and\ \bibinfo {author} {\bibfnamefont {D.}~\bibnamefont {Sen}},\ }\href
  {http://stacks.iop.org/0953-8984/27/i=27/a=275702} {\bibfield  {journal}
  {\bibinfo  {journal} {J. Phy. Condens. Matter}\ }\textbf {\bibinfo {volume}
  {{\bf 27}}},\ \bibinfo {pages} {275702} (\bibinfo {year} {2015})}\BibitemShut
  {NoStop}%
\bibitem [{\citenamefont {Dmytruk}\ \emph {et~al.}(2015)\citenamefont
  {Dmytruk}, \citenamefont {Trif},\ and\ \citenamefont {Simon}}]{Dmytruk:2015}%
  \BibitemOpen
  \bibfield  {author} {\bibinfo {author} {\bibfnamefont {O.}~\bibnamefont
  {Dmytruk}}, \bibinfo {author} {\bibfnamefont {M.}~\bibnamefont {Trif}}, \
  and\ \bibinfo {author} {\bibfnamefont {P.}~\bibnamefont {Simon}},\ }\href
  {https://link.aps.org/doi/10.1103/PhysRevB.92.245432} {\bibfield  {journal}
  {\bibinfo  {journal} {Phys. Rev. B}\ }\textbf {\bibinfo {volume} {{\bf
  92}}},\ \bibinfo {pages} {245432} (\bibinfo {year} {2015})}\BibitemShut
  {NoStop}%
\bibitem [{\citenamefont {Nadj-Perge}\ \emph {et~al.}(2014)\citenamefont
  {Nadj-Perge}, \citenamefont {Drozdov}, \citenamefont {Li}, \citenamefont
  {Chen}, \citenamefont {Jeon}, \citenamefont {Seo}, \citenamefont {MacDonald},
  \citenamefont {Bernevig},\ and\ \citenamefont {Yazdani}}]{NadjPerge:2014}%
  \BibitemOpen
  \bibfield  {author} {\bibinfo {author} {\bibfnamefont {S.}~\bibnamefont
  {Nadj-Perge}}, \bibinfo {author} {\bibfnamefont {I.~K.}\ \bibnamefont
  {Drozdov}}, \bibinfo {author} {\bibfnamefont {J.}~\bibnamefont {Li}},
  \bibinfo {author} {\bibfnamefont {H.}~\bibnamefont {Chen}}, \bibinfo {author}
  {\bibfnamefont {S.}~\bibnamefont {Jeon}}, \bibinfo {author} {\bibfnamefont
  {J.}~\bibnamefont {Seo}}, \bibinfo {author} {\bibfnamefont {A.~H.}\
  \bibnamefont {MacDonald}}, \bibinfo {author} {\bibfnamefont {B.~A.}\
  \bibnamefont {Bernevig}}, \ and\ \bibinfo {author} {\bibfnamefont
  {A.}~\bibnamefont {Yazdani}},\ }\href {\doibase 10.1126/science.1259327}
  {\bibfield  {journal} {\bibinfo  {journal} {Science}\ }\textbf {\bibinfo
  {volume} {{\bf 346}}},\ \bibinfo {pages} {602} (\bibinfo {year}
  {2014})}\BibitemShut {NoStop}%
\bibitem [{\citenamefont {Ruby}\ \emph {et~al.}(2015)\citenamefont {Ruby},
  \citenamefont {Pientka}, \citenamefont {Peng}, \citenamefont {von Oppen},
  \citenamefont {Heinrich},\ and\ \citenamefont {Franke}}]{Ruby:2015}%
  \BibitemOpen
  \bibfield  {author} {\bibinfo {author} {\bibfnamefont {M.}~\bibnamefont
  {Ruby}}, \bibinfo {author} {\bibfnamefont {F.}~\bibnamefont {Pientka}},
  \bibinfo {author} {\bibfnamefont {Y.}~\bibnamefont {Peng}}, \bibinfo {author}
  {\bibfnamefont {F.}~\bibnamefont {von Oppen}}, \bibinfo {author}
  {\bibfnamefont {B.~W.}\ \bibnamefont {Heinrich}}, \ and\ \bibinfo {author}
  {\bibfnamefont {K.~J.}\ \bibnamefont {Franke}},\ }\href
  {https://link.aps.org/doi/10.1103/PhysRevLett.115.197204} {\bibfield
  {journal} {\bibinfo  {journal} {Phys. Rev. Lett.}\ }\textbf {\bibinfo
  {volume} {{\bf 115}}},\ \bibinfo {pages} {197204} (\bibinfo {year}
  {2015})}\BibitemShut {NoStop}%
\bibitem [{\citenamefont {Pawlak}\ \emph {et~al.}(2016)\citenamefont {Pawlak},
  \citenamefont {Kisiel}, \citenamefont {Klinovaja}, \citenamefont {Meier},
  \citenamefont {Kawai}, \citenamefont {Glatzel}, \citenamefont {Loss},\ and\
  \citenamefont {Meyer}}]{Pawlak:2016}%
  \BibitemOpen
  \bibfield  {author} {\bibinfo {author} {\bibfnamefont {R.}~\bibnamefont
  {Pawlak}}, \bibinfo {author} {\bibfnamefont {M.}~\bibnamefont {Kisiel}},
  \bibinfo {author} {\bibfnamefont {J.}~\bibnamefont {Klinovaja}}, \bibinfo
  {author} {\bibfnamefont {T.}~\bibnamefont {Meier}}, \bibinfo {author}
  {\bibfnamefont {S.}~\bibnamefont {Kawai}}, \bibinfo {author} {\bibfnamefont
  {T.}~\bibnamefont {Glatzel}}, \bibinfo {author} {\bibfnamefont
  {D.}~\bibnamefont {Loss}}, \ and\ \bibinfo {author} {\bibfnamefont
  {E.}~\bibnamefont {Meyer}},\ }\href {http://dx.doi.org/10.1038/npjqi.2016.35}
  {\bibfield  {journal} {\bibinfo  {journal} {Npj Quantum Information}\
  }\textbf {\bibinfo {volume} {{\bf 2}}},\ \bibinfo {pages} {16035} (\bibinfo
  {year} {2016})}\BibitemShut {NoStop}%
\bibitem [{\citenamefont {Klinovaja}\ \emph {et~al.}(2013)\citenamefont
  {Klinovaja}, \citenamefont {Stano}, \citenamefont {Yazdani},\ and\
  \citenamefont {Loss}}]{Klinovaja:2013}%
  \BibitemOpen
  \bibfield  {author} {\bibinfo {author} {\bibfnamefont {J.}~\bibnamefont
  {Klinovaja}}, \bibinfo {author} {\bibfnamefont {P.}~\bibnamefont {Stano}},
  \bibinfo {author} {\bibfnamefont {A.}~\bibnamefont {Yazdani}}, \ and\
  \bibinfo {author} {\bibfnamefont {D.}~\bibnamefont {Loss}},\ }\href
  {http://link.aps.org/doi/10.1103/PhysRevLett.111.186805} {\bibfield
  {journal} {\bibinfo  {journal} {Phys. Rev. Lett.}\ }\textbf {\bibinfo
  {volume} {{\bf 111}}},\ \bibinfo {pages} {186805} (\bibinfo {year}
  {2013})}\BibitemShut {NoStop}%
\bibitem [{\citenamefont {Vazifeh}\ and\ \citenamefont
  {Franz}(2013)}]{Vazifeh:2013}%
  \BibitemOpen
  \bibfield  {author} {\bibinfo {author} {\bibfnamefont {M.~M.}\ \bibnamefont
  {Vazifeh}}\ and\ \bibinfo {author} {\bibfnamefont {M.}~\bibnamefont
  {Franz}},\ }\href {http://link.aps.org/doi/10.1103/PhysRevLett.111.206802}
  {\bibfield  {journal} {\bibinfo  {journal} {Phys. Rev. Lett.}\ }\textbf
  {\bibinfo {volume} {{\bf 111}}},\ \bibinfo {pages} {206802} (\bibinfo {year}
  {2013})}\BibitemShut {NoStop}%
\bibitem [{\citenamefont {Braunecker}\ and\ \citenamefont
  {Simon}(2013)}]{Braunecker:2013}%
  \BibitemOpen
  \bibfield  {author} {\bibinfo {author} {\bibfnamefont {B.}~\bibnamefont
  {Braunecker}}\ and\ \bibinfo {author} {\bibfnamefont {P.}~\bibnamefont
  {Simon}},\ }\href {http://link.aps.org/doi/10.1103/PhysRevLett.111.147202}
  {\bibfield  {journal} {\bibinfo  {journal} {Phys. Rev. Lett.}\ }\textbf
  {\bibinfo {volume} {{\bf 111}}},\ \bibinfo {pages} {147202} (\bibinfo {year}
  {2013})}\BibitemShut {NoStop}%
\bibitem [{\citenamefont {Nadj-Perge}\ \emph {et~al.}(2013)\citenamefont
  {Nadj-Perge}, \citenamefont {Drozdov}, \citenamefont {Bernevig},\ and\
  \citenamefont {Yazdani}}]{NadjPerge:2013}%
  \BibitemOpen
  \bibfield  {author} {\bibinfo {author} {\bibfnamefont {S.}~\bibnamefont
  {Nadj-Perge}}, \bibinfo {author} {\bibfnamefont {I.~K.}\ \bibnamefont
  {Drozdov}}, \bibinfo {author} {\bibfnamefont {B.~A.}\ \bibnamefont
  {Bernevig}}, \ and\ \bibinfo {author} {\bibfnamefont {A.}~\bibnamefont
  {Yazdani}},\ }\href {\doibase 10.1103/PhysRevB.88.020407} {\bibfield
  {journal} {\bibinfo  {journal} {Phys. Rev. B}\ }\textbf {\bibinfo {volume}
  {{\bf 88}}},\ \bibinfo {pages} {020407} (\bibinfo {year} {2013})}\BibitemShut
  {NoStop}%
\bibitem [{\citenamefont {Pientka}\ \emph {et~al.}(2013)\citenamefont
  {Pientka}, \citenamefont {Glazman},\ and\ \citenamefont {von
  Oppen}}]{Pientka:2013}%
  \BibitemOpen
  \bibfield  {author} {\bibinfo {author} {\bibfnamefont {F.}~\bibnamefont
  {Pientka}}, \bibinfo {author} {\bibfnamefont {L.~I.}\ \bibnamefont
  {Glazman}}, \ and\ \bibinfo {author} {\bibfnamefont {F.}~\bibnamefont {von
  Oppen}},\ }\href {http://link.aps.org/doi/10.1103/PhysRevB.88.155420}
  {\bibfield  {journal} {\bibinfo  {journal} {Phys. Rev. B}\ }\textbf {\bibinfo
  {volume} {{\bf 88}}},\ \bibinfo {pages} {155420} (\bibinfo {year}
  {2013})}\BibitemShut {NoStop}%
\bibitem [{\citenamefont {Awoga}\ \emph {et~al.}(2017)\citenamefont {Awoga},
  \citenamefont {Bj\"ornson},\ and\ \citenamefont
  {Black-Schaffer}}]{Awoga:2017}%
  \BibitemOpen
  \bibfield  {author} {\bibinfo {author} {\bibfnamefont {O.~A.}\ \bibnamefont
  {Awoga}}, \bibinfo {author} {\bibfnamefont {K.}~\bibnamefont {Bj\"ornson}}, \
  and\ \bibinfo {author} {\bibfnamefont {A.~M.}\ \bibnamefont
  {Black-Schaffer}},\ }\href
  {https://link.aps.org/doi/10.1103/PhysRevB.95.184511} {\bibfield  {journal}
  {\bibinfo  {journal} {Phys. Rev. B}\ }\textbf {\bibinfo {volume} {{\bf
  95}}},\ \bibinfo {pages} {184511} (\bibinfo {year} {2017})}\BibitemShut
  {NoStop}%
\bibitem [{\citenamefont {Mourik}\ \emph {et~al.}(2012)\citenamefont {Mourik},
  \citenamefont {Zuo}, \citenamefont {Frolov}, \citenamefont {Plissard},
  \citenamefont {Bakkers},\ and\ \citenamefont {Kouwenhoven}}]{Mourik:2012}%
  \BibitemOpen
  \bibfield  {author} {\bibinfo {author} {\bibfnamefont {V.}~\bibnamefont
  {Mourik}}, \bibinfo {author} {\bibfnamefont {K.}~\bibnamefont {Zuo}},
  \bibinfo {author} {\bibfnamefont {S.~M.}\ \bibnamefont {Frolov}}, \bibinfo
  {author} {\bibfnamefont {S.~R.}\ \bibnamefont {Plissard}}, \bibinfo {author}
  {\bibfnamefont {E.~P. A.~M.}\ \bibnamefont {Bakkers}}, \ and\ \bibinfo
  {author} {\bibfnamefont {L.~P.}\ \bibnamefont {Kouwenhoven}},\ }\href
  {\doibase 10.1126/science.1222360} {\bibfield  {journal} {\bibinfo  {journal}
  {Science}\ }\textbf {\bibinfo {volume} {{\bf 336}}},\ \bibinfo {pages} {1003}
  (\bibinfo {year} {2012})}\BibitemShut {NoStop}%
\bibitem [{\citenamefont {Deng}\ \emph {et~al.}(2012)\citenamefont {Deng},
  \citenamefont {Yu}, \citenamefont {Huang}, \citenamefont {Larsson},
  \citenamefont {Caroff},\ and\ \citenamefont {Xu}}]{Deng:2012}%
  \BibitemOpen
  \bibfield  {author} {\bibinfo {author} {\bibfnamefont {M.~T.}\ \bibnamefont
  {Deng}}, \bibinfo {author} {\bibfnamefont {C.~L.}\ \bibnamefont {Yu}},
  \bibinfo {author} {\bibfnamefont {G.~Y.}\ \bibnamefont {Huang}}, \bibinfo
  {author} {\bibfnamefont {M.}~\bibnamefont {Larsson}}, \bibinfo {author}
  {\bibfnamefont {P.}~\bibnamefont {Caroff}}, \ and\ \bibinfo {author}
  {\bibfnamefont {H.~Q.}\ \bibnamefont {Xu}},\ }\href {\doibase
  10.1021/nl303758w} {\bibfield  {journal} {\bibinfo  {journal} {Nano Letters}\
  }\textbf {\bibinfo {volume} {{\bf 12}}},\ \bibinfo {pages} {6414} (\bibinfo
  {year} {2012})}\BibitemShut {NoStop}%
\bibitem [{\citenamefont {Das}\ \emph {et~al.}(2012)\citenamefont {Das},
  \citenamefont {Ronen}, \citenamefont {Most}, \citenamefont {Oreg},
  \citenamefont {Heiblum},\ and\ \citenamefont {Shtrikman}}]{Das:2012}%
  \BibitemOpen
  \bibfield  {author} {\bibinfo {author} {\bibfnamefont {A.}~\bibnamefont
  {Das}}, \bibinfo {author} {\bibfnamefont {Y.}~\bibnamefont {Ronen}}, \bibinfo
  {author} {\bibfnamefont {Y.}~\bibnamefont {Most}}, \bibinfo {author}
  {\bibfnamefont {Y.}~\bibnamefont {Oreg}}, \bibinfo {author} {\bibfnamefont
  {M.}~\bibnamefont {Heiblum}}, \ and\ \bibinfo {author} {\bibfnamefont
  {H.}~\bibnamefont {Shtrikman}},\ }\href {http://dx.doi.org/10.1038/nphys2479}
  {\bibfield  {journal} {\bibinfo  {journal} {Nat. Phys.}\ }\textbf {\bibinfo
  {volume} {{\bf 8}}},\ \bibinfo {pages} {887} (\bibinfo {year}
  {2012})}\BibitemShut {NoStop}%
\bibitem [{\citenamefont {Churchill}\ \emph {et~al.}(2013)\citenamefont
  {Churchill}, \citenamefont {Fatemi}, \citenamefont {Grove-Rasmussen},
  \citenamefont {Deng}, \citenamefont {Caroff}, \citenamefont {Xu},\ and\
  \citenamefont {Marcus}}]{Churchill:2013}%
  \BibitemOpen
  \bibfield  {author} {\bibinfo {author} {\bibfnamefont {H.~O.~H.}\
  \bibnamefont {Churchill}}, \bibinfo {author} {\bibfnamefont {V.}~\bibnamefont
  {Fatemi}}, \bibinfo {author} {\bibfnamefont {K.}~\bibnamefont
  {Grove-Rasmussen}}, \bibinfo {author} {\bibfnamefont {M.~T.}\ \bibnamefont
  {Deng}}, \bibinfo {author} {\bibfnamefont {P.}~\bibnamefont {Caroff}},
  \bibinfo {author} {\bibfnamefont {H.~Q.}\ \bibnamefont {Xu}}, \ and\ \bibinfo
  {author} {\bibfnamefont {C.~M.}\ \bibnamefont {Marcus}},\ }\href
  {http://link.aps.org/doi/10.1103/PhysRevB.87.241401} {\bibfield  {journal}
  {\bibinfo  {journal} {Phys. Rev. B}\ }\textbf {\bibinfo {volume} {{\bf
  87}}},\ \bibinfo {pages} {241401} (\bibinfo {year} {2013})}\BibitemShut
  {NoStop}%
\bibitem [{\citenamefont {Finck}\ \emph {et~al.}(2013)\citenamefont {Finck},
  \citenamefont {Van~Harlingen}, \citenamefont {Mohseni}, \citenamefont
  {Jung},\ and\ \citenamefont {Li}}]{Finck:2013}%
  \BibitemOpen
  \bibfield  {author} {\bibinfo {author} {\bibfnamefont {A.~D.~K.}\
  \bibnamefont {Finck}}, \bibinfo {author} {\bibfnamefont {D.~J.}\ \bibnamefont
  {Van~Harlingen}}, \bibinfo {author} {\bibfnamefont {P.~K.}\ \bibnamefont
  {Mohseni}}, \bibinfo {author} {\bibfnamefont {K.}~\bibnamefont {Jung}}, \
  and\ \bibinfo {author} {\bibfnamefont {X.}~\bibnamefont {Li}},\ }\href
  {\doibase 10.1103/PhysRevLett.110.126406} {\bibfield  {journal} {\bibinfo
  {journal} {Phys. Rev. Lett.}\ }\textbf {\bibinfo {volume} {{\bf 110}}},\
  \bibinfo {pages} {126406} (\bibinfo {year} {2013})}\BibitemShut {NoStop}%
\bibitem [{\citenamefont {Chang}\ \emph {et~al.}(2015)\citenamefont {Chang},
  \citenamefont {Albrecht}, \citenamefont {Jespersen}, \citenamefont
  {Kuemmeth}, \citenamefont {Krogstrup}, \citenamefont {Nyg{\aa}rd},\ and\
  \citenamefont {Marcus}}]{Chang:2015}%
  \BibitemOpen
  \bibfield  {author} {\bibinfo {author} {\bibfnamefont {W.}~\bibnamefont
  {Chang}}, \bibinfo {author} {\bibfnamefont {S.~M.}\ \bibnamefont {Albrecht}},
  \bibinfo {author} {\bibfnamefont {T.~S.}\ \bibnamefont {Jespersen}}, \bibinfo
  {author} {\bibfnamefont {F.}~\bibnamefont {Kuemmeth}}, \bibinfo {author}
  {\bibfnamefont {P.}~\bibnamefont {Krogstrup}}, \bibinfo {author}
  {\bibfnamefont {J.}~\bibnamefont {Nyg{\aa}rd}}, \ and\ \bibinfo {author}
  {\bibfnamefont {C.~M.}\ \bibnamefont {Marcus}},\ }\href
  {http://dx.doi.org/10.1038/nnano.2014.306} {\bibfield  {journal} {\bibinfo
  {journal} {Nat. Nano.}\ }\textbf {\bibinfo {volume} {{\bf 10}}},\ \bibinfo
  {pages} {232} (\bibinfo {year} {2015})}\BibitemShut {NoStop}%
\bibitem [{\citenamefont {Albrecht}\ \emph {et~al.}(2016)\citenamefont
  {Albrecht}, \citenamefont {Higginbotham}, \citenamefont {Madsen},
  \citenamefont {Kuemmeth}, \citenamefont {Jespersen}, \citenamefont
  {Nyg{\aa}rd}, \citenamefont {Krogstrup},\ and\ \citenamefont
  {Marcus}}]{Albrecht:2016}%
  \BibitemOpen
  \bibfield  {author} {\bibinfo {author} {\bibfnamefont {S.~M.}\ \bibnamefont
  {Albrecht}}, \bibinfo {author} {\bibfnamefont {A.~P.}\ \bibnamefont
  {Higginbotham}}, \bibinfo {author} {\bibfnamefont {M.}~\bibnamefont
  {Madsen}}, \bibinfo {author} {\bibfnamefont {F.}~\bibnamefont {Kuemmeth}},
  \bibinfo {author} {\bibfnamefont {T.~S.}\ \bibnamefont {Jespersen}}, \bibinfo
  {author} {\bibfnamefont {J.}~\bibnamefont {Nyg{\aa}rd}}, \bibinfo {author}
  {\bibfnamefont {P.}~\bibnamefont {Krogstrup}}, \ and\ \bibinfo {author}
  {\bibfnamefont {C.~M.}\ \bibnamefont {Marcus}},\ }\href
  {http://dx.doi.org/10.1038/nature17162} {\bibfield  {journal} {\bibinfo
  {journal} {Nature}\ }\textbf {\bibinfo {volume} {{\bf 531}}},\ \bibinfo
  {pages} {206} (\bibinfo {year} {2016})}\BibitemShut {NoStop}%
\bibitem [{\citenamefont {Deng}\ \emph {et~al.}(2016)\citenamefont {Deng},
  \citenamefont {Vaitiekenas}, \citenamefont {Hansen}, \citenamefont {Danon},
  \citenamefont {Leijnse}, \citenamefont {Flensberg}, \citenamefont
  {Nyg{\aa}rd}, \citenamefont {Krogstrup},\ and\ \citenamefont
  {Marcus}}]{Deng:2016}%
  \BibitemOpen
  \bibfield  {author} {\bibinfo {author} {\bibfnamefont {M.~T.}\ \bibnamefont
  {Deng}}, \bibinfo {author} {\bibfnamefont {S.}~\bibnamefont {Vaitiekenas}},
  \bibinfo {author} {\bibfnamefont {E.~B.}\ \bibnamefont {Hansen}}, \bibinfo
  {author} {\bibfnamefont {J.}~\bibnamefont {Danon}}, \bibinfo {author}
  {\bibfnamefont {M.}~\bibnamefont {Leijnse}}, \bibinfo {author} {\bibfnamefont
  {K.}~\bibnamefont {Flensberg}}, \bibinfo {author} {\bibfnamefont
  {J.}~\bibnamefont {Nyg{\aa}rd}}, \bibinfo {author} {\bibfnamefont
  {P.}~\bibnamefont {Krogstrup}}, \ and\ \bibinfo {author} {\bibfnamefont
  {C.~M.}\ \bibnamefont {Marcus}},\ }\href
  {http://science.sciencemag.org/content/354/6319/1557} {\bibfield  {journal}
  {\bibinfo  {journal} {Science}\ }\textbf {\bibinfo {volume} {{\bf 354}}},\
  \bibinfo {pages} {1557} (\bibinfo {year} {2016})}\BibitemShut {NoStop}%
\bibitem [{\citenamefont {Gazibegovic}\ \emph {et~al.}(2017)\citenamefont
  {Gazibegovic}, \citenamefont {Car}, \citenamefont {Zhang}, \citenamefont
  {Balk}, \citenamefont {Logan}, \citenamefont {de~Moor}, \citenamefont
  {Cassidy}, \citenamefont {Schmits}, \citenamefont {Xu}, \citenamefont {Wang},
  \citenamefont {Krogstrup}, \citenamefont {Op~het Veld}, \citenamefont {Zuo},
  \citenamefont {Vos}, \citenamefont {Shen}, \citenamefont {Bouman},
  \citenamefont {Shojaei}, \citenamefont {Pennachio}, \citenamefont {Lee},
  \citenamefont {van Veldhoven}, \citenamefont {Koelling}, \citenamefont
  {Verheijen}, \citenamefont {Kouwenhoven}, \citenamefont {Palmstr{\o}m},\ and\
  \citenamefont {Bakkers}}]{Gazibegovic:2017}%
  \BibitemOpen
  \bibfield  {author} {\bibinfo {author} {\bibfnamefont {S.}~\bibnamefont
  {Gazibegovic}}, \bibinfo {author} {\bibfnamefont {D.}~\bibnamefont {Car}},
  \bibinfo {author} {\bibfnamefont {H.}~\bibnamefont {Zhang}}, \bibinfo
  {author} {\bibfnamefont {S.~C.}\ \bibnamefont {Balk}}, \bibinfo {author}
  {\bibfnamefont {J.~A.}\ \bibnamefont {Logan}}, \bibinfo {author}
  {\bibfnamefont {M.~W.~A.}\ \bibnamefont {de~Moor}}, \bibinfo {author}
  {\bibfnamefont {M.~C.}\ \bibnamefont {Cassidy}}, \bibinfo {author}
  {\bibfnamefont {R.}~\bibnamefont {Schmits}}, \bibinfo {author} {\bibfnamefont
  {D.}~\bibnamefont {Xu}}, \bibinfo {author} {\bibfnamefont {G.}~\bibnamefont
  {Wang}}, \bibinfo {author} {\bibfnamefont {P.}~\bibnamefont {Krogstrup}},
  \bibinfo {author} {\bibfnamefont {R.~L.~M.}\ \bibnamefont {Op~het Veld}},
  \bibinfo {author} {\bibfnamefont {K.}~\bibnamefont {Zuo}}, \bibinfo {author}
  {\bibfnamefont {Y.}~\bibnamefont {Vos}}, \bibinfo {author} {\bibfnamefont
  {J.}~\bibnamefont {Shen}}, \bibinfo {author} {\bibfnamefont {D.}~\bibnamefont
  {Bouman}}, \bibinfo {author} {\bibfnamefont {B.}~\bibnamefont {Shojaei}},
  \bibinfo {author} {\bibfnamefont {D.}~\bibnamefont {Pennachio}}, \bibinfo
  {author} {\bibfnamefont {J.~S.}\ \bibnamefont {Lee}}, \bibinfo {author}
  {\bibfnamefont {P.~J.}\ \bibnamefont {van Veldhoven}}, \bibinfo {author}
  {\bibfnamefont {S.}~\bibnamefont {Koelling}}, \bibinfo {author}
  {\bibfnamefont {M.~A.}\ \bibnamefont {Verheijen}}, \bibinfo {author}
  {\bibfnamefont {L.~P.}\ \bibnamefont {Kouwenhoven}}, \bibinfo {author}
  {\bibfnamefont {C.~J.}\ \bibnamefont {Palmstr{\o}m}}, \ and\ \bibinfo
  {author} {\bibfnamefont {E.~P. A.~M.}\ \bibnamefont {Bakkers}},\ }\href
  {http://dx.doi.org/10.1038/nature23468} {\bibfield  {journal} {\bibinfo
  {journal} {Nature}\ }\textbf {\bibinfo {volume} {{\bf 548}}},\ \bibinfo
  {pages} {434} (\bibinfo {year} {2017})}\BibitemShut {NoStop}%
\bibitem [{\citenamefont {Zhang}\ \emph {et~al.}(2018)\citenamefont {Zhang},
  \citenamefont {Liu}, \citenamefont {Gazibegovic}, \citenamefont {Xu},
  \citenamefont {Logan}, \citenamefont {Wang}, \citenamefont {van Loo},
  \citenamefont {Bommer}, \citenamefont {de~Moor}, \citenamefont {Car},
  \citenamefont {Op~het Veld}, \citenamefont {van Veldhoven}, \citenamefont
  {Koelling}, \citenamefont {Verheijen}, \citenamefont {Pendharkar},
  \citenamefont {Pennachio}, \citenamefont {Shojaei}, \citenamefont {Lee},
  \citenamefont {Palmstr{\o}m}, \citenamefont {Bakkers}, \citenamefont
  {Sarma},\ and\ \citenamefont {Kouwenhoven}}]{Zhang:2017_2}%
  \BibitemOpen
  \bibfield  {author} {\bibinfo {author} {\bibfnamefont {H.}~\bibnamefont
  {Zhang}}, \bibinfo {author} {\bibfnamefont {C.-X.}\ \bibnamefont {Liu}},
  \bibinfo {author} {\bibfnamefont {S.}~\bibnamefont {Gazibegovic}}, \bibinfo
  {author} {\bibfnamefont {D.}~\bibnamefont {Xu}}, \bibinfo {author}
  {\bibfnamefont {J.~A.}\ \bibnamefont {Logan}}, \bibinfo {author}
  {\bibfnamefont {G.}~\bibnamefont {Wang}}, \bibinfo {author} {\bibfnamefont
  {N.}~\bibnamefont {van Loo}}, \bibinfo {author} {\bibfnamefont {J.~D.~S.}\
  \bibnamefont {Bommer}}, \bibinfo {author} {\bibfnamefont {M.~W.~A.}\
  \bibnamefont {de~Moor}}, \bibinfo {author} {\bibfnamefont {D.}~\bibnamefont
  {Car}}, \bibinfo {author} {\bibfnamefont {R.~L.~M.}\ \bibnamefont {Op~het
  Veld}}, \bibinfo {author} {\bibfnamefont {P.~J.}\ \bibnamefont {van
  Veldhoven}}, \bibinfo {author} {\bibfnamefont {S.}~\bibnamefont {Koelling}},
  \bibinfo {author} {\bibfnamefont {M.~A.}\ \bibnamefont {Verheijen}}, \bibinfo
  {author} {\bibfnamefont {M.}~\bibnamefont {Pendharkar}}, \bibinfo {author}
  {\bibfnamefont {D.~J.}\ \bibnamefont {Pennachio}}, \bibinfo {author}
  {\bibfnamefont {B.}~\bibnamefont {Shojaei}}, \bibinfo {author} {\bibfnamefont
  {J.~S.}\ \bibnamefont {Lee}}, \bibinfo {author} {\bibfnamefont {C.~J.}\
  \bibnamefont {Palmstr{\o}m}}, \bibinfo {author} {\bibfnamefont {E.~P. A.~M.}\
  \bibnamefont {Bakkers}}, \bibinfo {author} {\bibfnamefont {S.~D.}\
  \bibnamefont {Sarma}}, \ and\ \bibinfo {author} {\bibfnamefont {L.~P.}\
  \bibnamefont {Kouwenhoven}},\ }\href
  {https://www.nature.com/articles/nature26142} {\bibfield  {journal} {\bibinfo
   {journal} {Nature}\ }\textbf {\bibinfo {volume} {{\bf 556}}},\ \bibinfo
  {pages} {74} (\bibinfo {year} {2018})}\BibitemShut {NoStop}%
\bibitem [{\citenamefont {Kjaergaard}\ \emph {et~al.}(2016)\citenamefont
  {Kjaergaard}, \citenamefont {Nichele}, \citenamefont {Suominen},
  \citenamefont {Nowak}, \citenamefont {Wimmer}, \citenamefont {Akhmerov},
  \citenamefont {Folk}, \citenamefont {Flensberg}, \citenamefont {Shabani},
  \citenamefont {Palmstr{\o}m},\ and\ \citenamefont
  {Marcus}}]{Kjaergaard:2016}%
  \BibitemOpen
  \bibfield  {author} {\bibinfo {author} {\bibfnamefont {M.}~\bibnamefont
  {Kjaergaard}}, \bibinfo {author} {\bibfnamefont {F.}~\bibnamefont {Nichele}},
  \bibinfo {author} {\bibfnamefont {H.~J.}\ \bibnamefont {Suominen}}, \bibinfo
  {author} {\bibfnamefont {M.~P.}\ \bibnamefont {Nowak}}, \bibinfo {author}
  {\bibfnamefont {M.}~\bibnamefont {Wimmer}}, \bibinfo {author} {\bibfnamefont
  {A.~R.}\ \bibnamefont {Akhmerov}}, \bibinfo {author} {\bibfnamefont {J.~A.}\
  \bibnamefont {Folk}}, \bibinfo {author} {\bibfnamefont {K.}~\bibnamefont
  {Flensberg}}, \bibinfo {author} {\bibfnamefont {J.}~\bibnamefont {Shabani}},
  \bibinfo {author} {\bibfnamefont {C.~J.}\ \bibnamefont {Palmstr{\o}m}}, \
  and\ \bibinfo {author} {\bibfnamefont {C.~M.}\ \bibnamefont {Marcus}},\
  }\href {http://dx.doi.org/10.1038/ncomms12841} {\bibfield  {journal}
  {\bibinfo  {journal} {Nat. Commun.}\ }\textbf {\bibinfo {volume} {{\bf 7}}},\
  \bibinfo {pages} {12841} (\bibinfo {year} {2016})}\BibitemShut {NoStop}%
\bibitem [{\citenamefont {Shabani}\ \emph {et~al.}(2016)\citenamefont
  {Shabani}, \citenamefont {Kjaergaard}, \citenamefont {Suominen},
  \citenamefont {Kim}, \citenamefont {Nichele}, \citenamefont {Pakrouski},
  \citenamefont {Stankevic}, \citenamefont {Lutchyn}, \citenamefont
  {Krogstrup}, \citenamefont {Feidenhans'l}, \citenamefont {Kraemer},
  \citenamefont {Nayak}, \citenamefont {Troyer}, \citenamefont {Marcus},\ and\
  \citenamefont {Palmstr\o{}m}}]{Shabani:2016}%
  \BibitemOpen
  \bibfield  {author} {\bibinfo {author} {\bibfnamefont {J.}~\bibnamefont
  {Shabani}}, \bibinfo {author} {\bibfnamefont {M.}~\bibnamefont {Kjaergaard}},
  \bibinfo {author} {\bibfnamefont {H.~J.}\ \bibnamefont {Suominen}}, \bibinfo
  {author} {\bibfnamefont {Y.}~\bibnamefont {Kim}}, \bibinfo {author}
  {\bibfnamefont {F.}~\bibnamefont {Nichele}}, \bibinfo {author} {\bibfnamefont
  {K.}~\bibnamefont {Pakrouski}}, \bibinfo {author} {\bibfnamefont
  {T.}~\bibnamefont {Stankevic}}, \bibinfo {author} {\bibfnamefont {R.~M.}\
  \bibnamefont {Lutchyn}}, \bibinfo {author} {\bibfnamefont {P.}~\bibnamefont
  {Krogstrup}}, \bibinfo {author} {\bibfnamefont {R.}~\bibnamefont
  {Feidenhans'l}}, \bibinfo {author} {\bibfnamefont {S.}~\bibnamefont
  {Kraemer}}, \bibinfo {author} {\bibfnamefont {C.}~\bibnamefont {Nayak}},
  \bibinfo {author} {\bibfnamefont {M.}~\bibnamefont {Troyer}}, \bibinfo
  {author} {\bibfnamefont {C.~M.}\ \bibnamefont {Marcus}}, \ and\ \bibinfo
  {author} {\bibfnamefont {C.~J.}\ \bibnamefont {Palmstr\o{}m}},\ }\href
  {https://link.aps.org/doi/10.1103/PhysRevB.93.155402} {\bibfield  {journal}
  {\bibinfo  {journal} {Phys. Rev. B}\ }\textbf {\bibinfo {volume} {{\bf
  93}}},\ \bibinfo {pages} {155402} (\bibinfo {year} {2016})}\BibitemShut
  {NoStop}%
\bibitem [{\citenamefont {Kjaergaard}\ \emph {et~al.}(2017)\citenamefont
  {Kjaergaard}, \citenamefont {Suominen}, \citenamefont {Nowak}, \citenamefont
  {Akhmerov}, \citenamefont {Shabani}, \citenamefont {Palmstr\o{}m},
  \citenamefont {Nichele},\ and\ \citenamefont {Marcus}}]{Kjaergaard:2017}%
  \BibitemOpen
  \bibfield  {author} {\bibinfo {author} {\bibfnamefont {M.}~\bibnamefont
  {Kjaergaard}}, \bibinfo {author} {\bibfnamefont {H.~J.}\ \bibnamefont
  {Suominen}}, \bibinfo {author} {\bibfnamefont {M.~P.}\ \bibnamefont {Nowak}},
  \bibinfo {author} {\bibfnamefont {A.~R.}\ \bibnamefont {Akhmerov}}, \bibinfo
  {author} {\bibfnamefont {J.}~\bibnamefont {Shabani}}, \bibinfo {author}
  {\bibfnamefont {C.~J.}\ \bibnamefont {Palmstr\o{}m}}, \bibinfo {author}
  {\bibfnamefont {F.}~\bibnamefont {Nichele}}, \ and\ \bibinfo {author}
  {\bibfnamefont {C.~M.}\ \bibnamefont {Marcus}},\ }\href
  {https://link.aps.org/doi/10.1103/PhysRevApplied.7.034029} {\bibfield
  {journal} {\bibinfo  {journal} {Phys. Rev. Applied}\ }\textbf {\bibinfo
  {volume} {{\bf 7}}},\ \bibinfo {pages} {034029} (\bibinfo {year}
  {2017})}\BibitemShut {NoStop}%
\bibitem [{\citenamefont {Suominen}\ \emph {et~al.}(2017)\citenamefont
  {Suominen}, \citenamefont {Kjaergaard}, \citenamefont {Hamilton},
  \citenamefont {Shabani}, \citenamefont {Palmstr\o{}m}, \citenamefont
  {Marcus},\ and\ \citenamefont {Nichele}}]{Suominen:2017}%
  \BibitemOpen
  \bibfield  {author} {\bibinfo {author} {\bibfnamefont {H.~J.}\ \bibnamefont
  {Suominen}}, \bibinfo {author} {\bibfnamefont {M.}~\bibnamefont
  {Kjaergaard}}, \bibinfo {author} {\bibfnamefont {A.~R.}\ \bibnamefont
  {Hamilton}}, \bibinfo {author} {\bibfnamefont {J.}~\bibnamefont {Shabani}},
  \bibinfo {author} {\bibfnamefont {C.~J.}\ \bibnamefont {Palmstr\o{}m}},
  \bibinfo {author} {\bibfnamefont {C.~M.}\ \bibnamefont {Marcus}}, \ and\
  \bibinfo {author} {\bibfnamefont {F.}~\bibnamefont {Nichele}},\ }\href
  {https://link.aps.org/doi/10.1103/PhysRevLett.119.176805} {\bibfield
  {journal} {\bibinfo  {journal} {Phys. Rev. Lett.}\ }\textbf {\bibinfo
  {volume} {{\bf 119}}},\ \bibinfo {pages} {176805} (\bibinfo {year}
  {2017})}\BibitemShut {NoStop}%
\bibitem [{\citenamefont {Nichele}\ \emph {et~al.}(2017)\citenamefont
  {Nichele}, \citenamefont {Drachmann}, \citenamefont {Whiticar}, \citenamefont
  {O'Farrell}, \citenamefont {Suominen}, \citenamefont {Fornieri},
  \citenamefont {Wang}, \citenamefont {Gardner}, \citenamefont {Thomas},
  \citenamefont {Hatke}, \citenamefont {Krogstrup}, \citenamefont {Manfra},
  \citenamefont {Flensberg},\ and\ \citenamefont {Marcus}}]{Nichele:2017}%
  \BibitemOpen
  \bibfield  {author} {\bibinfo {author} {\bibfnamefont {F.}~\bibnamefont
  {Nichele}}, \bibinfo {author} {\bibfnamefont {A.~C.~C.}\ \bibnamefont
  {Drachmann}}, \bibinfo {author} {\bibfnamefont {A.~M.}\ \bibnamefont
  {Whiticar}}, \bibinfo {author} {\bibfnamefont {E.~C.~T.}\ \bibnamefont
  {O'Farrell}}, \bibinfo {author} {\bibfnamefont {H.~J.}\ \bibnamefont
  {Suominen}}, \bibinfo {author} {\bibfnamefont {A.}~\bibnamefont {Fornieri}},
  \bibinfo {author} {\bibfnamefont {T.}~\bibnamefont {Wang}}, \bibinfo {author}
  {\bibfnamefont {G.~C.}\ \bibnamefont {Gardner}}, \bibinfo {author}
  {\bibfnamefont {C.}~\bibnamefont {Thomas}}, \bibinfo {author} {\bibfnamefont
  {A.~T.}\ \bibnamefont {Hatke}}, \bibinfo {author} {\bibfnamefont
  {P.}~\bibnamefont {Krogstrup}}, \bibinfo {author} {\bibfnamefont {M.~J.}\
  \bibnamefont {Manfra}}, \bibinfo {author} {\bibfnamefont {K.}~\bibnamefont
  {Flensberg}}, \ and\ \bibinfo {author} {\bibfnamefont {C.~M.}\ \bibnamefont
  {Marcus}},\ }\href {https://link.aps.org/doi/10.1103/PhysRevLett.119.136803}
  {\bibfield  {journal} {\bibinfo  {journal} {Phys. Rev. Lett.}\ }\textbf
  {\bibinfo {volume} {{\bf 119}}},\ \bibinfo {pages} {136803} (\bibinfo {year}
  {2017})}\BibitemShut {NoStop}%
\bibitem [{\citenamefont {Reeg}\ \emph
  {et~al.}(2017{\natexlab{a}})\citenamefont {Reeg}, \citenamefont {Loss},\ and\
  \citenamefont {Klinovaja}}]{Reeg:2017_3}%
  \BibitemOpen
  \bibfield  {author} {\bibinfo {author} {\bibfnamefont {C.}~\bibnamefont
  {Reeg}}, \bibinfo {author} {\bibfnamefont {D.}~\bibnamefont {Loss}}, \ and\
  \bibinfo {author} {\bibfnamefont {J.}~\bibnamefont {Klinovaja}},\ }\href
  {https://link.aps.org/doi/10.1103/PhysRevB.96.125426} {\bibfield  {journal}
  {\bibinfo  {journal} {Phys. Rev. B}\ }\textbf {\bibinfo {volume} {{\bf
  96}}},\ \bibinfo {pages} {125426} (\bibinfo {year}
  {2017}{\natexlab{a}})}\BibitemShut {NoStop}%
\bibitem [{\citenamefont {Reeg}\ \emph {et~al.}(2018)\citenamefont {Reeg},
  \citenamefont {Loss},\ and\ \citenamefont {Klinovaja}}]{Reeg:2018}%
  \BibitemOpen
  \bibfield  {author} {\bibinfo {author} {\bibfnamefont {C.}~\bibnamefont
  {Reeg}}, \bibinfo {author} {\bibfnamefont {D.}~\bibnamefont {Loss}}, \ and\
  \bibinfo {author} {\bibfnamefont {J.}~\bibnamefont {Klinovaja}},\ }\href
  {https://link.aps.org/doi/10.1103/PhysRevB.97.165425} {\bibfield  {journal}
  {\bibinfo  {journal} {Phys. Rev. B}\ }\textbf {\bibinfo {volume} {{\bf
  97}}},\ \bibinfo {pages} {165425} (\bibinfo {year} {2018})}\BibitemShut
  {NoStop}%
\bibitem [{\citenamefont {Volkov}\ \emph {et~al.}(1995)\citenamefont {Volkov},
  \citenamefont {Magn{\'e}e}, \citenamefont {van Wees},\ and\ \citenamefont
  {Klapwijk}}]{Volkov:1995}%
  \BibitemOpen
  \bibfield  {author} {\bibinfo {author} {\bibfnamefont {A.~F.}\ \bibnamefont
  {Volkov}}, \bibinfo {author} {\bibfnamefont {P.~H.~C.}\ \bibnamefont
  {Magn{\'e}e}}, \bibinfo {author} {\bibfnamefont {B.~J.}\ \bibnamefont {van
  Wees}}, \ and\ \bibinfo {author} {\bibfnamefont {T.~M.}\ \bibnamefont
  {Klapwijk}},\ }\href {\doibase
  http://dx.doi.org/10.1016/0921-4534(94)02429-4} {\bibfield  {journal}
  {\bibinfo  {journal} {Physica C}\ }\textbf {\bibinfo {volume} {{\bf 242}}},\
  \bibinfo {pages} {261 } (\bibinfo {year} {1995})}\BibitemShut {NoStop}%
\bibitem [{\citenamefont {Fagas}\ \emph {et~al.}(2005)\citenamefont {Fagas},
  \citenamefont {Tkachov}, \citenamefont {Pfund},\ and\ \citenamefont
  {Richter}}]{Fagas:2005}%
  \BibitemOpen
  \bibfield  {author} {\bibinfo {author} {\bibfnamefont {G.}~\bibnamefont
  {Fagas}}, \bibinfo {author} {\bibfnamefont {G.}~\bibnamefont {Tkachov}},
  \bibinfo {author} {\bibfnamefont {A.}~\bibnamefont {Pfund}}, \ and\ \bibinfo
  {author} {\bibfnamefont {K.}~\bibnamefont {Richter}},\ }\href {\doibase
  10.1103/PhysRevB.71.224510} {\bibfield  {journal} {\bibinfo  {journal} {Phys.
  Rev. B}\ }\textbf {\bibinfo {volume} {{\bf 71}}},\ \bibinfo {pages} {224510}
  (\bibinfo {year} {2005})}\BibitemShut {NoStop}%
\bibitem [{\citenamefont {Tkachov}(2005)}]{Tkachov:2005}%
  \BibitemOpen
  \bibfield  {author} {\bibinfo {author} {\bibfnamefont {G.}~\bibnamefont
  {Tkachov}},\ }\href {\doibase http://dx.doi.org/10.1016/j.physc.2004.10.015}
  {\bibfield  {journal} {\bibinfo  {journal} {Physica C}\ }\textbf {\bibinfo
  {volume} {{\bf 417}}},\ \bibinfo {pages} {127 } (\bibinfo {year}
  {2005})}\BibitemShut {NoStop}%
\bibitem [{\citenamefont {Reeg}\ and\ \citenamefont
  {Maslov}(2016)}]{Reeg:2016}%
  \BibitemOpen
  \bibfield  {author} {\bibinfo {author} {\bibfnamefont {C.~R.}\ \bibnamefont
  {Reeg}}\ and\ \bibinfo {author} {\bibfnamefont {D.~L.}\ \bibnamefont
  {Maslov}},\ }\href {http://link.aps.org/doi/10.1103/PhysRevB.94.020501}
  {\bibfield  {journal} {\bibinfo  {journal} {Phys. Rev. B}\ }\textbf {\bibinfo
  {volume} {{\bf 94}}},\ \bibinfo {pages} {020501} (\bibinfo {year}
  {2016})}\BibitemShut {NoStop}%
\bibitem [{\citenamefont {Reeg}\ \emph
  {et~al.}(2017{\natexlab{b}})\citenamefont {Reeg}, \citenamefont {Klinovaja},\
  and\ \citenamefont {Loss}}]{Reeg:2017}%
  \BibitemOpen
  \bibfield  {author} {\bibinfo {author} {\bibfnamefont {C.}~\bibnamefont
  {Reeg}}, \bibinfo {author} {\bibfnamefont {J.}~\bibnamefont {Klinovaja}}, \
  and\ \bibinfo {author} {\bibfnamefont {D.}~\bibnamefont {Loss}},\ }\href
  {https://link.aps.org/doi/10.1103/PhysRevB.96.081301} {\bibfield  {journal}
  {\bibinfo  {journal} {Phys. Rev. B}\ }\textbf {\bibinfo {volume} {{\bf
  96}}},\ \bibinfo {pages} {081301} (\bibinfo {year}
  {2017}{\natexlab{b}})}\BibitemShut {NoStop}%
\bibitem [{\citenamefont {Sau}\ \emph {et~al.}(2010{\natexlab{b}})\citenamefont
  {Sau}, \citenamefont {Lutchyn}, \citenamefont {Tewari},\ and\ \citenamefont
  {Das~Sarma}}]{Sau:2010prox}%
  \BibitemOpen
  \bibfield  {author} {\bibinfo {author} {\bibfnamefont {J.~D.}\ \bibnamefont
  {Sau}}, \bibinfo {author} {\bibfnamefont {R.~M.}\ \bibnamefont {Lutchyn}},
  \bibinfo {author} {\bibfnamefont {S.}~\bibnamefont {Tewari}}, \ and\ \bibinfo
  {author} {\bibfnamefont {S.}~\bibnamefont {Das~Sarma}},\ }\href
  {http://link.aps.org/doi/10.1103/PhysRevB.82.094522} {\bibfield  {journal}
  {\bibinfo  {journal} {Phys. Rev. B}\ }\textbf {\bibinfo {volume} {{\bf
  82}}},\ \bibinfo {pages} {094522} (\bibinfo {year}
  {2010}{\natexlab{b}})}\BibitemShut {NoStop}%
\bibitem [{\citenamefont {Potter}\ and\ \citenamefont
  {Lee}(2011)}]{Potter:2011}%
  \BibitemOpen
  \bibfield  {author} {\bibinfo {author} {\bibfnamefont {A.~C.}\ \bibnamefont
  {Potter}}\ and\ \bibinfo {author} {\bibfnamefont {P.~A.}\ \bibnamefont
  {Lee}},\ }\href {\doibase 10.1103/PhysRevB.83.184520} {\bibfield  {journal}
  {\bibinfo  {journal} {Phys. Rev. B}\ }\textbf {\bibinfo {volume} {{\bf
  83}}},\ \bibinfo {pages} {184520} (\bibinfo {year} {2011})}\BibitemShut
  {NoStop}%
\bibitem [{\citenamefont {Kopnin}\ and\ \citenamefont
  {Melnikov}(2011)}]{Kopnin:2011}%
  \BibitemOpen
  \bibfield  {author} {\bibinfo {author} {\bibfnamefont {N.~B.}\ \bibnamefont
  {Kopnin}}\ and\ \bibinfo {author} {\bibfnamefont {A.~S.}\ \bibnamefont
  {Melnikov}},\ }\href {\doibase 10.1103/PhysRevB.84.064524} {\bibfield
  {journal} {\bibinfo  {journal} {Phys. Rev. B}\ }\textbf {\bibinfo {volume}
  {{\bf 84}}},\ \bibinfo {pages} {064524} (\bibinfo {year} {2011})}\BibitemShut
  {NoStop}%
\bibitem [{\citenamefont {Zyuzin}\ \emph {et~al.}(2013)\citenamefont {Zyuzin},
  \citenamefont {Rainis}, \citenamefont {Klinovaja},\ and\ \citenamefont
  {Loss}}]{Zyuzin:2013}%
  \BibitemOpen
  \bibfield  {author} {\bibinfo {author} {\bibfnamefont {A.~A.}\ \bibnamefont
  {Zyuzin}}, \bibinfo {author} {\bibfnamefont {D.}~\bibnamefont {Rainis}},
  \bibinfo {author} {\bibfnamefont {J.}~\bibnamefont {Klinovaja}}, \ and\
  \bibinfo {author} {\bibfnamefont {D.}~\bibnamefont {Loss}},\ }\href
  {http://link.aps.org/doi/10.1103/PhysRevLett.111.056802} {\bibfield
  {journal} {\bibinfo  {journal} {Phys. Rev. Lett.}\ }\textbf {\bibinfo
  {volume} {{\bf 111}}},\ \bibinfo {pages} {056802} (\bibinfo {year}
  {2013})}\BibitemShut {NoStop}%
\bibitem [{\citenamefont {van Heck}\ \emph {et~al.}(2016)\citenamefont {van
  Heck}, \citenamefont {Lutchyn},\ and\ \citenamefont
  {Glazman}}]{vanHeck:2016}%
  \BibitemOpen
  \bibfield  {author} {\bibinfo {author} {\bibfnamefont {B.}~\bibnamefont {van
  Heck}}, \bibinfo {author} {\bibfnamefont {R.~M.}\ \bibnamefont {Lutchyn}}, \
  and\ \bibinfo {author} {\bibfnamefont {L.~I.}\ \bibnamefont {Glazman}},\
  }\href {http://link.aps.org/doi/10.1103/PhysRevB.93.235431} {\bibfield
  {journal} {\bibinfo  {journal} {Phys. Rev. B}\ }\textbf {\bibinfo {volume}
  {{\bf 93}}},\ \bibinfo {pages} {235431} (\bibinfo {year} {2016})}\BibitemShut
  {NoStop}%
\bibitem [{\citenamefont {Reeg}\ and\ \citenamefont
  {Maslov}(2017)}]{Reeg:2017_2}%
  \BibitemOpen
  \bibfield  {author} {\bibinfo {author} {\bibfnamefont {C.}~\bibnamefont
  {Reeg}}\ and\ \bibinfo {author} {\bibfnamefont {D.~L.}\ \bibnamefont
  {Maslov}},\ }\href {https://link.aps.org/doi/10.1103/PhysRevB.95.205439}
  {\bibfield  {journal} {\bibinfo  {journal} {Phys. Rev. B}\ }\textbf {\bibinfo
  {volume} {{\bf 95}}},\ \bibinfo {pages} {205439} (\bibinfo {year}
  {2017})}\BibitemShut {NoStop}%
\end{thebibliography}%

\end{document}